\documentclass[conference]{IEEEtran}

\usepackage{bibentry}
\usepackage{booktabs}
\usepackage{cite}
\usepackage{color}

\usepackage{enumitem}
\usepackage{float}
\usepackage{flushend}
\usepackage{graphicx}
\usepackage{grffile}
\usepackage{multirow}
\usepackage[caption=false,font=footnotesize]{subfig}
\usepackage{tabularx}
\usepackage{textcomp}
\usepackage{times}
\usepackage{url}
\usepackage{xspace}

\DeclareUnicodeCharacter{2009}{ }

\begin{document}

\newif\ifdraft

\ifdraft
  \newcommand{\note}[1]{{\textcolor{blue}{ ***Note: #1 }}\xspace}
  \newcommand{\jhanote}[1]{{\textcolor{red}{ ***Shantenu: #1 }}\xspace}
  \newcommand{\mtnote}[1]{{\textcolor{orange}{ ***Matteo: #1 }}\xspace}
  \newcommand{\amnote}[1]{{\textcolor{green}{ ***Andre: #1 }}\xspace}
\else
  \newcommand{\note}[1]{}
  \newcommand{\jhanote}[1]{}
  \newcommand{\mtnote}[1]{}
  \newcommand{\amnote}[1]{}
\fi

\newcommand{\project}{IMPECCABLE }
\newcommand{\esmacscg}{ESMACS--CG }
\newcommand{\wfcg}{WF3--CG }
\newcommand{\esmacsfg}{ESMACS--FG }
\newcommand{\wffg}{WF3--FG }
\newcommand{\up}{\vspace*{-0.5em}}

\title{Scalable HPC \& AI Infrastructure for COVID-19 Therapeutics}

\author{
  \IEEEauthorblockN{
    Hyungro Lee\IEEEauthorrefmark{1},
    Andre Merzky\IEEEauthorrefmark{1},
    Li Tan\IEEEauthorrefmark{5},
    Mikhail Titov\IEEEauthorrefmark{1},
    Matteo Turilli\IEEEauthorrefmark{1}}
  \IEEEauthorblockN{
    Dario~Alfe\IEEEauthorrefmark{2}\,\IEEEauthorrefmark{7},
    Agastya Bhati\IEEEauthorrefmark{2},
    Alex Brace\IEEEauthorrefmark{4},
    Austin Clyde\IEEEauthorrefmark{2},
    Peter Coveney\IEEEauthorrefmark{2}\,\IEEEauthorrefmark{6},
    Heng~Ma\IEEEauthorrefmark{4},\\
    Arvind Ramanathan\IEEEauthorrefmark{4},
    Rick Stevens\IEEEauthorrefmark{3}\,\IEEEauthorrefmark{4},
    Anda Trifan\IEEEauthorrefmark{4}\,\IEEEauthorrefmark{8},
    Hubertus Van Dam\IEEEauthorrefmark{5},\\
    Shunzhou Wan\IEEEauthorrefmark{2},
    Sean Wilkinson\IEEEauthorrefmark{9},
    Shantenu Jha\IEEEauthorrefmark{1}\,\IEEEauthorrefmark{5}}
  \IEEEauthorblockA{
    \IEEEauthorrefmark{1}Rutgers University,
    \IEEEauthorrefmark{5}Brookhaven National Laboratory}
  \IEEEauthorblockA{
    \IEEEauthorrefmark{2}University College London,
    \IEEEauthorrefmark{3}University of Chicago,
    \IEEEauthorrefmark{4}Argonne National Laboratory,
    \IEEEauthorrefmark{6}University of Amsterdam,\\
    \IEEEauthorrefmark{7}University of Naples Federico II,
    \IEEEauthorrefmark{8}University of Illinois Urbana-Champaign,
    \IEEEauthorrefmark{9}Oak Ridge National Laboratory}}

\maketitle


\begin{abstract}
COVID-19 has claimed more $10^{6}$ lives and resulted in over $40\times10^{6}$
infections. There is an urgent need to identify drugs that can inhibit
SARS-CoV-2. In response, the DOE recently established  the Medical
Therapeutics project as part of the National Virtual Biotechnology Laboratory,
and tasked it with creating the computational infrastructure and methods
necessary to advance therapeutics development. We discuss innovations in
computational infrastructure and methods that are accelerating and advancing
drug design. Specifically, we describe several methods that integrate
artificial intelligence and simulation-based approaches, and the design of
computational infrastructure to support these methods at scale. We discuss
their implementation and characterize their performance, and highlight science
advances that these capabilities have enabled.
\end{abstract}

\section{Introduction}
\label{sec:intro}

Considering the universe of about 10$^{68}$ possible compounds to traverse for
effective drugs, there is an immediate need for more efficient and more
effective  frameworks for early stage drug
discovery~\cite{Bohacek_et_al:2010}. {\it In silico} drug discovery is a
promising but computationally intensive and complex approach. There is a
critical need to improve {\it in silico} methodologies to accelerate and
select better lead compounds that can proceed to later stages of the drug
discovery protocol at the best of times; it is nothing short of a societal and
intellectual grand challenge in the age of COVID-19.




A fundamental challenge for {\it in silico} drug discovery
is the need to
cover multiple physical length- and timescales, while spanning an enormous
combinatorial and chemical space. In this complex landscape, no
single methodological approach can achieve the necessary accuracy of lead
compound selection with required computational efficiency. Multiple
methodological innovations that accelerate lead compound selections are
needed. These methods must in turn, be able to utilize advanced and scalable
computational infrastructure.




The primary objective of this paper is to describe the scalable computational
infrastructure developed to support innovative methods across diverse and
heterogeneous platforms. It discusses four distinct workflows, their
computational characteristics, and their implementation using abstractions and
RADICAL-Cybertools middleware building blocks. It discusses diverse types of
computational capabilities required, the sustained and high watermark
performance and scale.




These capabilities provide the computational fabric of the US-DOE National
Virtual Biotechnology Laboratory in combination with resources from the EU
Centre of Excellence in Computational Biomedicine. To provide a sense of the
impact and scale of operations: these methods and infrastructure  are being
used to screen over 4.2 billion molecules against over a dozen drug targets in
SARS-CoV-2. So far, over a 1000 compounds have been identified and
experimentally validated, resulting in advanced testing for dozens of hits.
The campaign used several million node-hours across diverse HPC platforms,
including TACC Frontera, Livermore Computing Lassen, ANL Theta (and
associated A100 nodes), LRZ SuperMUC-NG, and ORNL Summit to obtain
scientific results.

This paper is organized as follows: In \S2 we outline the computational
campaign, and describe the constituent scientific methods and their
computational properties. Put together, the collective impact of these methods
on drug discovery process is greater than the sum of the individual parts. In
\S3 we discuss core middleware building blocks used to develop the
computational infrastructure. We describe the design and implementation to
support diverse workflows that comprise the campaign over multiple
heterogeneous HPC platforms. The following section discusses the performance
characteristics and highlights the  necessary extensions to overcome
challenges of scale and system-software fragility. We conclude in \S5 with
diverse measures of scientific impact towards therapeutic advances by
highlighting science results emanating from using these methods and
infrastructure for a mission-oriented sustained computational campaign.

\section{Computational Campaign}
\label{sec:cc}
\begin{figure}[h]
\centering
\includegraphics[width=\linewidth]{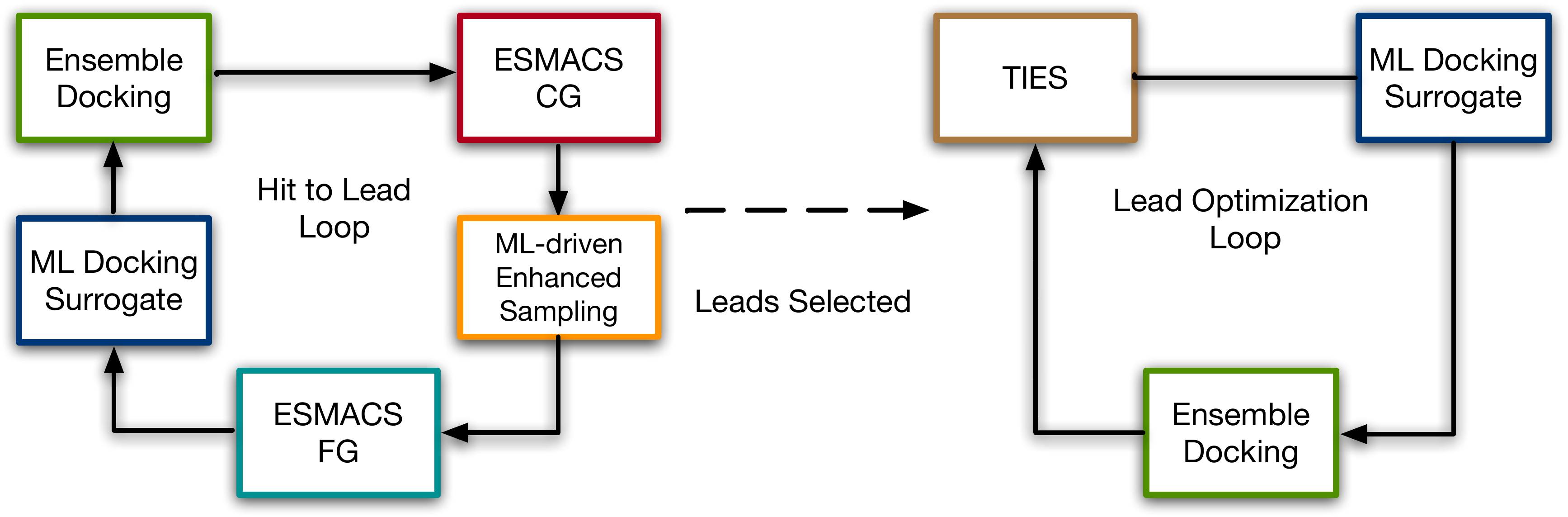}
\up\up\up
\caption{The computational campaign to advance COVID-19 therapeutics has two
coupled loops: drug candidates go through four stages in the Hit-to-Lead loop;
a small set of  drugs are selected for the Lead Optimization loop. We focus on
the following methods: Ensemble Docking, both coarse-grained (CG) and
fine-grained (FG) ESMACS, ML-driven Enhanced Sampling and TIES.}
\up\up
\label{fig:cc}
\end{figure}

The campaign for discovering new `hits' (i.e., viable drug-like small
molecules), and optimizing these hits to viable lead molecules (i.e., that
show potential to inhibit viral activity) consists of two iterative loops
(Fig. \ref{fig:cc}). In ensemble docking programs (Sec.~\ref{sec:Docking}),
small molecules are `docked' against ensembles of conformational states
determined from a particular COVID-19 protein target. This is followed by a
coarse-grained free-energy estimation (ESMACS-CG) to determine if the docked
molecule and the protein target of interest can indeed interact
(Sec.~\ref{sec:H2LGen}).
\esmacscg samples limited relevant conformational states, outputs from
\esmacscg into ML-driven enhanced sampling methods
(Sec.~\ref{sssection:mlsampling}), are taken through a further fine-grained
refinement of the binding free-energy (ESMACS-FG; Sec.~\ref{sec:H2LGen}). The
outputs from \esmacsfg are fed into a machine learning (ML) surrogate for
docking, which allows us to quickly estimate whether a given molecule can
indeed bind to the COVID-19 protein target. Simultaneously, a secondary
iterative loop is developed for a subset of compounds that show promising
results from ESMACS-CG, where certain functional groups of promising hits are
optimized for protein-target interactions using TIES \textemdash{} a method of
lead optimization (Sec.~\ref{ssection:leadopt}).

\subsection{Computational Protein-Ligand Docking}
\label{sec:Docking}

Protein-ligand docking is a computational pipeline of ligand 3D structure
(conformer) enumeration, exhaustive docking and scoring, and pose scoring. The
input requires a protein structure with a designed binding region, or a
crystallized ligand from which a region can be inferred, as well as a database
of molecules to dock in SMILES format \textemdash{} a compact representation
of a 2D molecule.

To take the 2D structures to 3D structures ready for structural docking,
proteinization and conformer generation is performed using Omega--Tautomers
and, if stereochemistry is not specified, enantiomers are enumerated prior to
conformer generation~\cite{oetoolkit}. Typically, tautomers and enantiomers are
enumerated for the incoming proposed analog or perturbation to the previous
ligand. Conformer generation is performed on the ensemble of structures,
generating 200-800 3D conformers for every enantiomer and reasonable tautomer
generated. Once the set of 3D structures are enumerated from the 2D smiles,
each one is docked against the pocket and scored. The best scoring pose is
returned along with its ChemGauss4 score from exhaustive rigid
docking~\cite{mcgann2003gaussian}.

\subsection{ML-Driven Enhanced Sampling}
\label{sssection:mlsampling}

Machine learning tools quantify statistical insights into the
time-dependent structural changes a biomolecule undergoes in
simulations~\cite{Maisuradze_2008}, identify events that characterize
large-scale conformational changes at multiple timescales, build
low-dimensional representations of simulation data capturing biophysical or
biochemical information, use these low-dimensional representations to infer
kinetically and energetically coherent conformational sub-states, and obtain
quantitative comparisons with experiments.

Recently, we developed convolutional variational autoencoders (CVAE) that
automatically reduce the high dimensionality of MD trajectories and cluster
conformations into a small number of conformational states that share similar
structural, and energetic characteristics~\cite{Bhowmik_2018}. We apply these
approaches on the ESMACS and TIES simulations, outlined shortly. We also used
CVAE to drive adaptive simulations for protein folding, and demonstrated that
adaptive sampling techniques can provide at least an order of magnitude
speedup~\cite{lee2019deepdrivemd}. These approaches provide acceleration of
``rare'' events \textemdash{} important to study protein-ligand interactions,
while leveraging supercomputing platforms~\cite{lee2019deepdrivemd}.


\subsection{Hit-to-Lead Optimization}
\label{sec:H2LGen}

Hit-to-Lead (H2L) is a step in the drug discovery process where promising lead
compounds are identified from initial hits generated at preceding stages. It
involves evaluation of initial hits followed by some optimization of
potentially good compounds to achieve nanomolar affinities. The change in free
energy between free and bound states of protein and ligand, also known as
binding affinity, is a promising measure of the binding potency of a molecule,
and is used as a parameter for evaluating and optimizing hits at H2L
stage.

We employ the enhanced sampling of molecular dynamics with approximation of
continuum solvent (ESMACS)~\cite{brd4} protocol, for
estimating binding affinities of protein-ligand complexes. It involves
performing an ensemble of molecular dynamics (MD) simulations followed by free
energy estimation on the conformations so generated using a semi-empirical
method called molecular mechanics Poisson-Boltzmann Surface Area
(MMPBSA). The free energies so calculated for the ensemble of
conformations are analyzed in a statistically robust manner yielding precise
free energy predictions for any given complex. This is particularly important
given the fact that the usual practice of performing MMPBSA calculations on
conformations generated using a single MD simulation does not give reliable
binding affinities. Consequently, ESMACS predictions can be used to rank a
large number of hits based on their binding affinities. ESMACS is able to
handle large variations in ligand structures, and hence is very suitable for
H2L stage where hits have been picked out after covering a substantial region
of chemical space.
The information and data generated with ESMACS can also be used to train a ML
algorithm to improve its predictive capability.

\subsection{Lead Optimization}
\label{ssection:leadopt}

Lead Optimization (LO) is the final step of pre-clinical drug discovery process.
It involves altering the structures of selected lead compounds in order to
improve their properties such as selectivity, potency and pharmacokinetic
parameters. Binding affinity is a useful parameter to make \textit{in silico}
predictions about effects of any chemical alteration in a lead molecule.
However, LO requires theoretically more accurate (without much/any
approximations) methods for predictions with high confidence. In addition,
relative binding affinity of pairs of compounds which are structurally similar
are of interest, rendering ESMACS unsuitable for LO. We employ thermodynamic
integration with enhanced sampling (TIES)~\cite{ties}, which is based on an
alchemical free energy method called thermodynamic integration (TI)~\cite{ti1}
which fulfill conditions for LO. Alchemical methods involve calculating free
energy along a non-physical thermodynamic pathway to get relative free energy
between the two end-points. Usually, the alchemical pathway corresponds to
transformation of one chemical species into another defined with a coupling
parameter ($\lambda$), ranging between 0 and 1. TIES involves performing an
ensemble simulation at each $\lambda$ value to generate the ensemble of
conformations to be used for calculating relative free energy. It also involves
performing a robust error analysis to yield relative binding affinities with
statistically meaningful error bars. The parameters such as the size of the
ensemble and the length of simulations are determined keeping in mind the
desired level of precision in the results~\cite{ties}.

\section{Computational Infrastructure}
\label{sec:infrastructure}
We use the term ``task'' to indicate a stand-alone process that has well-defined
input, output, termination criteria, and dedicated resources. For example, a
task can indicate an executable which performs a simulation or a data processing
analysis, executing on one or more nodes. A workflow is
comprised of tasks with dependencies, whereas a workload represents a set of
tasks without dependences or whose dependencies have been resolved. Thus, the
tasks of a workload could, resources permitting, be executed concurrently.

\subsection{RADICAL-Cybertools Overview}


RADICAL-Cybertools (RCT) are software systems developed to support the execution
of heterogeneous workflows and workloads on one or more high-performance
computing (HPC) infrastructures. RCT have three main components: RADICAL-SAGA
(RS)~\cite{merzky2015saga}, RADICAL-Pilot (RP)~\cite{merzky2018using} and
RADICAL-Ensemble Toolkit (EnTK)~\cite{balasubramanian2018harnessing}.

RS is a Python implementation of the Open Grid Forum SAGA standard GFD.90, a
high-level interface to distributed infrastructure components like job
schedulers, file transfer and resource provisioning services. RS enables
interoperability across heterogeneous distributed infrastructures.


RP is a Python implementation of the pilot paradigm and architectural
pattern~\cite{turilli2018comprehensive}. Pilot systems enable users to submit
pilot jobs to computing infrastructures and then use the resources acquired by
the pilot to execute one or more workloads. Tasks are executed concurrently and
sequentially, depending on the available resources. RP can execute single or
multi core tasks within a single compute node, or across multiple nodes. RP
isolates the execution of each tasks into a dedicated process, enabling
concurrent execution of heterogeneous tasks. RP offers three unique features
compared to other pilot systems on HPC systems: (1) concurrent execution of
heterogeneous tasks on the same pilot; (2) support of all the major HPC batch
systems; and, (3) support of more than twelve methods to launch tasks.

EnTK is a Python implementation of a workflow engine, designed to support the
programming of applications comprised of ensembles of tasks. EnTK executes tasks
concurrently or sequentially, depending on their arbitrary priority relation.
Tasks are grouped into stages and stages into pipelines depending on the
priority relation among tasks. Tasks without reciprocal priority relations can
be grouped into the same stage; tasks that need to be executed before other
tasks have to be grouped into different stages. Stages are grouped into
pipelines and, in turn, multiple pipelines can be executed either concurrently
or sequentially. EnTK uses RP, allowing the execution of workflows with
heterogeneous tasks.

\subsection{Supporting Multiple Task Execution Modes}

Workflows 1--4 (\S\ref{sec:cc}) have different tasks types and performance
requirements that, in turn, require different execution approaches. We discuss
three pilot-based task execution frameworks that we developed to support the
execution of workflows 1--4. We provide a brief comparison of the three
approaches.

\begin{figure*}
  \up\up
  \centering
  \subfloat[][]{
      \includegraphics[width=0.31\textwidth]{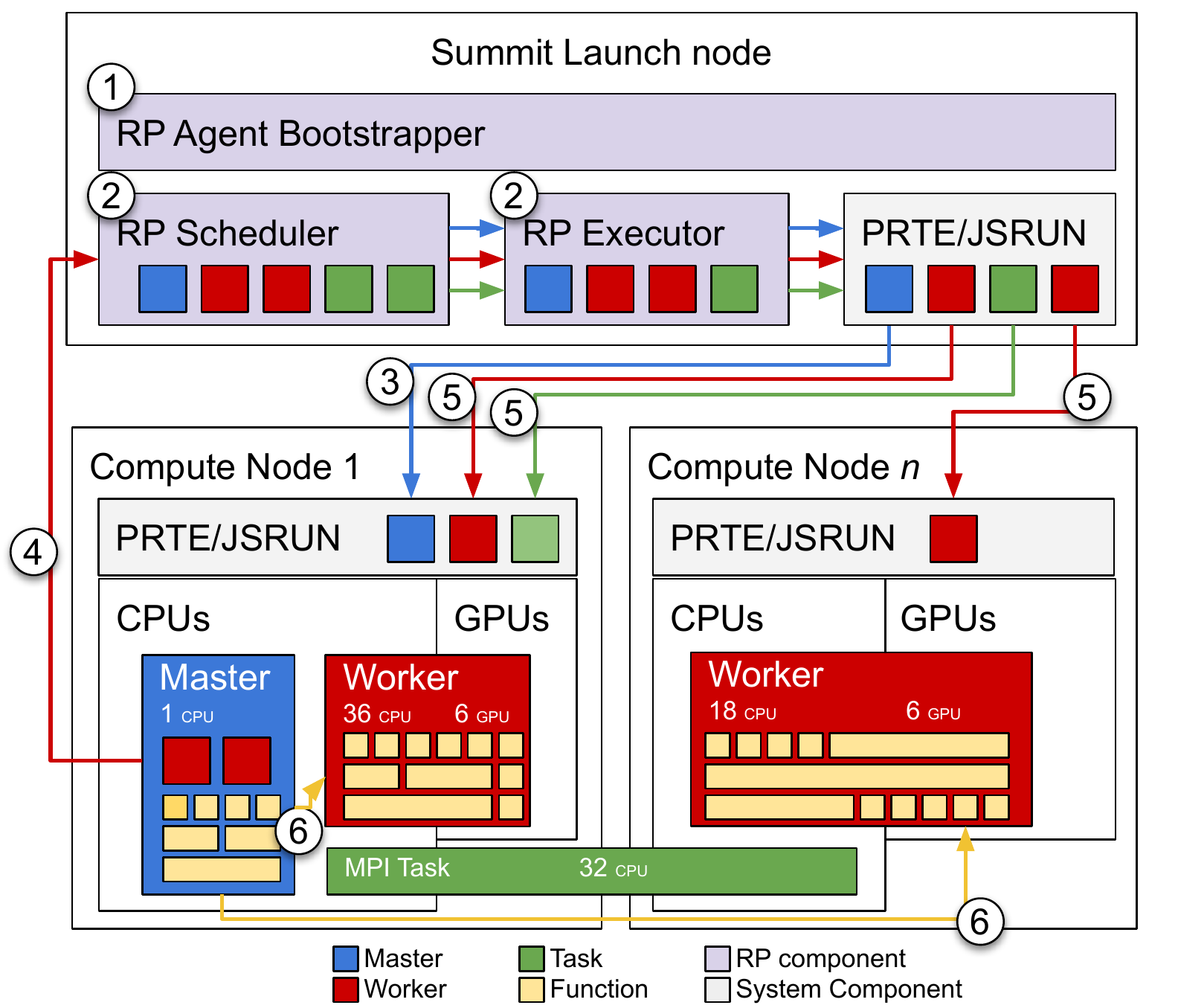}
      \label{sfig:raptor}}
  \hfill
  \subfloat[][]{
      \includegraphics[width=0.31\textwidth]{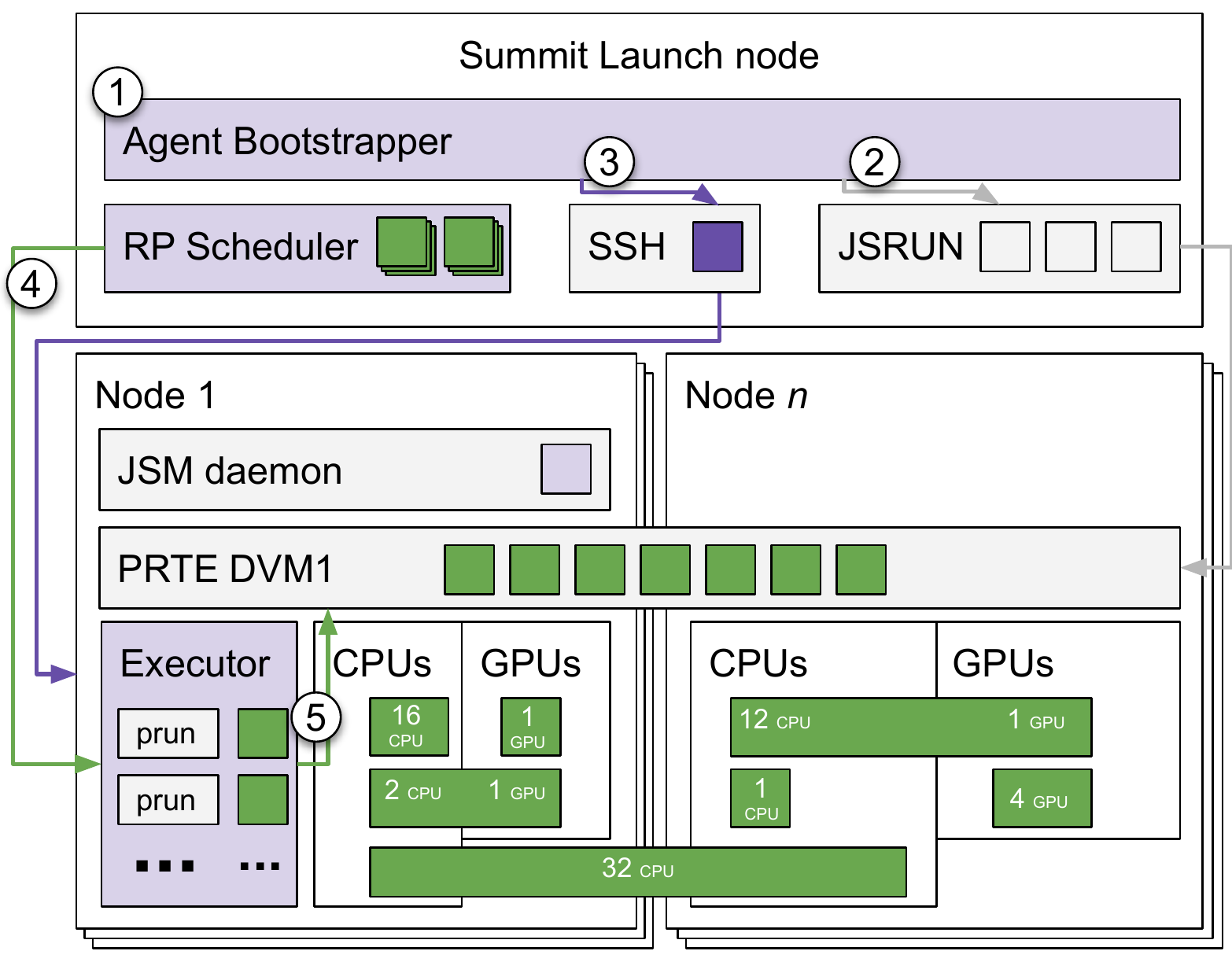}
      \label{sfig:multidvm}}
  \hfill
  \subfloat[][]{
      \includegraphics[width=0.31\textwidth]{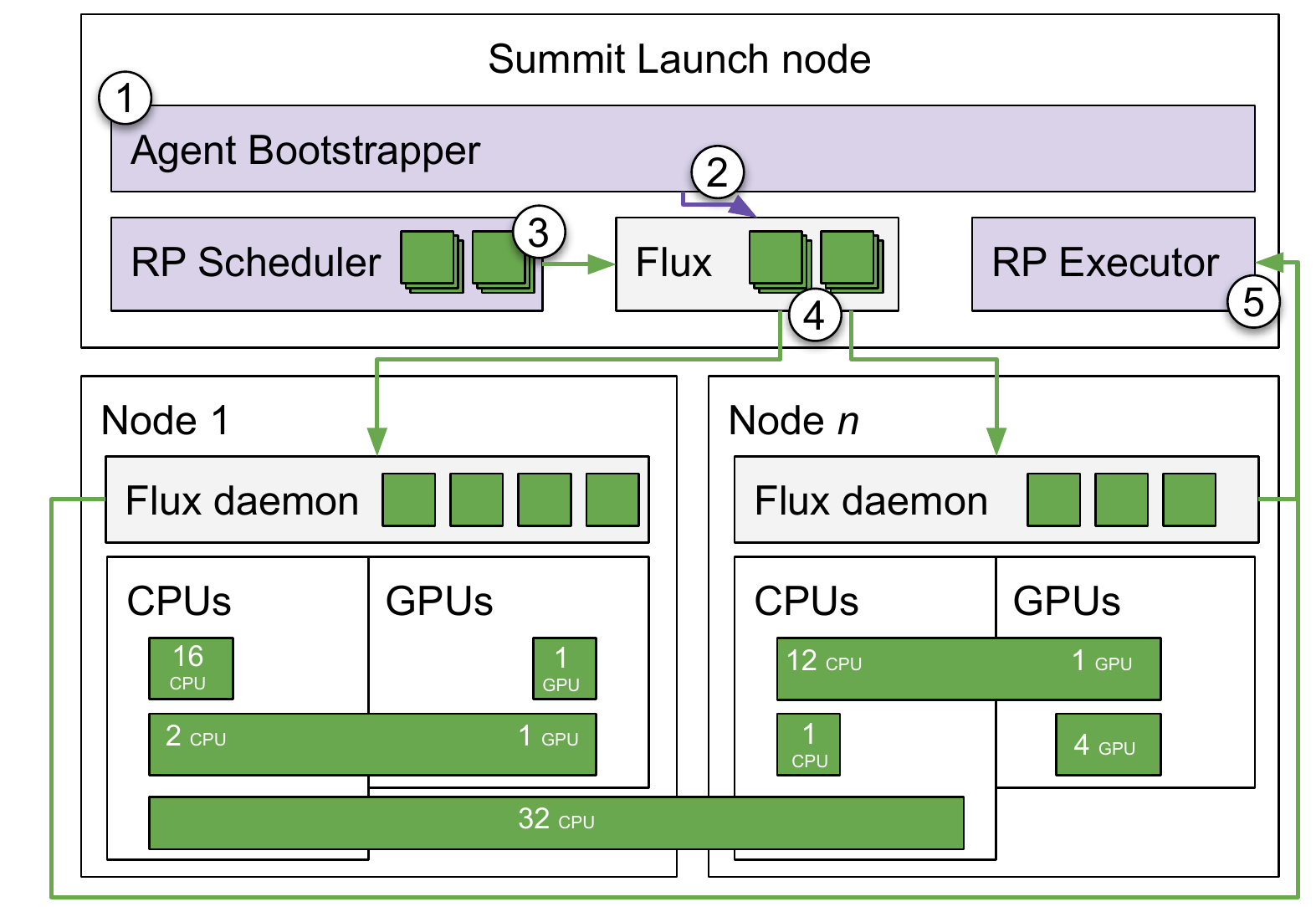}
      \label{sfig:flux}}
  \up
  \caption{Pilot-based task execution frameworks implemented using RADIAL-Pilot.}
  \up\up
  \label{fig:lms}
\end{figure*}

\subsubsection{Execution Mode I: RAPTOR}
\label{sssection:raptor}

RP can execute a special type of task, called ``worker'', that can interpret and
execute Python functions. We used this feature to implement a RP-based
master/worker framework called RAPTOR (RAdical-Pilot Task OveRlay), to
distribute multiple Python functions across multiple workers. RAPTOR enables
parallel execution of those functions while RP implements capabilities to code
both master and worker tasks, and to schedule their execution on the HPC
resources acquired by submitting a job.

Fig.~\ref{sfig:raptor} illustrates the implementation of RAPTOR on Summit. Once
RP has acquired its resources by submiting a job to Summit's batch system, RP
bootstraps its Agent (Fig.~\ref{sfig:raptor}-1) and launches a task
scheduler and a task executor (Fig.~\ref{sfig:raptor}-2). RP Scheduler and RP
Executor schedule and launch one or more masters on one the compute nodes
(Fig.~\ref{sfig:raptor}-3) via either JSRUN~\cite{quintero2019ibm} or
PRRTE~\cite{castain2018pmix}. Once running, each master schedules one or more
workers on RP Scheduler (Fig.~\ref{sfig:raptor}-4). Those workers are then
launched on more compute nodes by RP Executor (Fig.~\ref{sfig:raptor}-5).
Finally, each master schedules function calls on the available workers for
execution (Fig.~\ref{sfig:raptor}-6), load-balancing across workers to obtain
maximal resource utilization. The only change needed to use RAPTOR on other
machines is a switch of the launch method for the master and worker tasks, e.g.,
on Frontera, from JSRUN to srun.

\subsubsection{Execution Mode II: Using multi-DVM}
\label{sssection:multidvm}

RP supports diverse task launch methods, depending on the availability of
specific software systems on the target resources. On Summit, Frontera and
Lassen at the Lawrence Livermore National Laboratory, RP supports the use of
the Process Management Interface for Exascale~(PMIx) and the PMIx Reference
RunTime Environment (PRRTE)~\cite{castain2018pmix}. PMIx is an open source
standard that provides methods to interact with system-level resource managers
and process launch mechanisms. PRRTE provides a portable runtime layer that
users can leverage to launch a PMIx server. PRRTE includes a persistent mode
called Distributed Virtual Machine (DVM), which uses system-native launch
mechanisms to bootstrap an overlay runtime environment, which
can be used to launch tasks via the PMIx interface.

One advantage of using PRRTE/PMIx to place and launch stand-alone tasks on
thousands of compute nodes, is that they allow for using multiple concurrent
DVMs. This enables partitioning of the task execution over multiple,
independent sub-systems, reducing the communication and coordination pressure
on each sub-system. This improves performance and resilience to PRRTE/PMIx
implementation fragility.


Fig.~\ref{sfig:multidvm} shows the integration between RP and PRRTE/PMIx on
Summit. RP bootstraps its Agent (Fig.~\ref{sfig:multidvm}-1) and, different from
the RAPTOR implementation described in Fig.~\ref{sfig:raptor}, Agent launches a
set of DVMs, each spanning multiple compute nodes
(Fig.~\ref{sfig:multidvm}-2). Agent also uses \texttt{ssh} to execute one or
more RP Executor on one or more compute nodes (Fig.~\ref{sfig:multidvm}-3). Once
the DVMs and executors become available, RP schedules tasks on each executor
(Fig.~\ref{sfig:multidvm}-4). Each executor then uses one or more DVMs to place
and then launch those tasks (Fig.~\ref{sfig:multidvm}-5).

\subsubsection{Execution Mode III: Flux}
\label{sssection:flux}

PRRTE/PMIx introduce a variety of overheads~\cite{turilli2019characterizing})
and their current implementations are still fragile, especially when
scheduling more than 20,000 tasks on more than 32 DVMs. Overheads and
fragility lead to low resource utilization and unrecoverable failures. For
these reasons, RP also supports the use of Flux~\cite{ahn2014flux} as an
alternative system to schedule, place and launch tasks implemented as
stand-alone processes.
Fig.~\ref{sfig:flux} illustrates the integration between RP and Flux. After
bootstrapping (Fig.~\ref{sfig:flux}-1), RP launches Flux
(Fig.~\ref{sfig:flux}-2) and schedules tasks on it for execution
(Fig.~\ref{sfig:flux}-3). Flux schedules, places and launches tasks on Summit's
compute nodes via its daemons (Fig.~\ref{sfig:flux}-4). RP Executor keeps track
of task completion (Fig.~\ref{sfig:flux}-5), and communicates this information
to RP Scheduler, based upon which RP Scheduler passes more tasks to Flux for
execution.

\subsection{DeepDriveMD}
\label{ssection:DeepDriveMD}

To support the requirements of ML-driven enhanced sampling
(\S\ref{sssection:mlsampling}), we developed
DeepDriveMD~\cite{lee2019deepdrivemd} to employ deep learning techniques,
pre-trained models and tuned hyperparameters in conjunction with molecular
dynamics (MD) simulations for adaptive sampling. Specifically, DeepDriveMD
couples a deep learning (DL) network --- called convolutional variational
autoencoder (CVAE)) --- to multiple MD simulations, to cluster MD trajectories
into a small number of conformational states. Insights gained from clustering is
used to steer the ensemble MD simulations. This may include either starting new
simulations (i.e., expanding the pool of initial MD simulations), or killing
unproductive MD simulations (i.e., simulations stuck in metastable states).

DeepDriveMD supports the following computational approach: (1) use an ensemble
of MD simulations to generate initial MD data; (2) a `training' run consisting of
a ML algorithm; (3) an `inference' step where novel starting points for MD are
identified; and (4) spawn new MD simulations. DeepDriveMD is built upon
EnTK, uses RP for advanced resource management, and is extensible to other
learning methods and models, as well as other MD coupling schemes. The current
implementation of DeepDriveMD utilizes Tensorflow/Keras (with Horovod for
distributed data parallel training) and PyTorch. Typically, a run of DeepDriveMD
requires 20 nodes on Summit.

\subsection{Heterogeneous Task Placement}
\label{ssection:infwfhybrid}

Depending on the task launch method, RP places tasks on specific compute
nodes, cores and GPU (Figs.~\ref{sfig:raptor} and~\ref{sfig:multidvm}). This
placement allows for efficient scheduling of tasks on heterogeneous resources.
When scheduling tasks that require different amounts of cores and/or GPUs, RP
keeps tracks of the available slots on each compute node of its pilot.
Depending on availability, RP schedules MPI tasks within and across compute
nodes and reserves a CPU core for each GPU task.

Currently, RP supports four scheduling algorithms: continuous, torus, noop and
flux. Continuous is a general purpose algorithm that enables task
ordering, task colocation on the same or on different nodes,
based on arrival order or explicit task tagging. Torus is a special-purpose
algorithm written to support BlueGene architectures, noop allows to pass single
or bulk tasks keeping track only of their execution state, and flux delegates
scheduling to the Flux framework.

RP opens a large optimization space for specific scheduling algorithms. For
example, our continuous scheduler prioritizes tasks that require large amount of
cores/GPU so to maximize resource utilization. This could be further extended
with explicit clustering or by including information about the execution time of
each task.

These capabilities are used to concurrently execute WF3 and 4
(\S\ref{sec:H2LGen} and \S\ref{ssection:leadopt}), reducing time-to-solution and
improving resource utilization at scale. WF3 and WF4 capture ESMACS and TIES
respectively, which are both MD-based protocols to compute binding free
energies. Both protocols involve multiple stages of equilibration and MD
simulations of protein-ligand complexes. Specifically, ESMACS protocol uses the
OpenMM MD engine on GPUs, while the TIES protocol uses NAMD on CPUs. Leveraging
RP's capabilities, we merge these two ``workflows'' into an integrated hybrid
workflow with heterogeneous tasks which utilize CPU and GPU concurrently.

\begin{figure}[h!]
  \vspace{-0.5in}
  \up
   \begin{center}
    \includegraphics[width=0.5\textwidth]{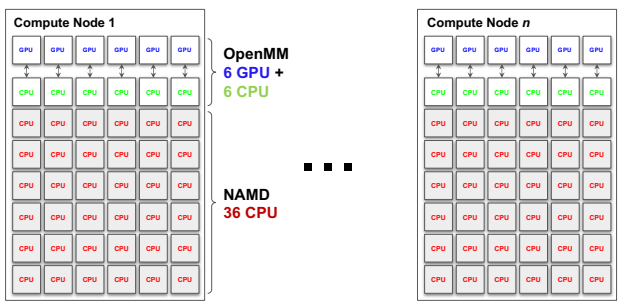}
  \end{center}
  \vspace{-0.75in}
  \caption{Using RP's heterogeneous task placement: WF3 is comprised of OpenMM
           which runs on GPUs; WF4 uses NAMD which runs on CPUs. A hybrid
           workflow combines WF3 and WF4, concurrently using the CPUs and GPUs
           of the same Summit node. This increases node utilization
           significantly.}
  \up\up
  \label{fig:hybrid}
\end{figure}

Fig.\ref{fig:hybrid} is a schematic where OpenMM simulations are tasks placed on
GPUs, while NAMD simulations are MPI multicore tasks on GPUs. Given that one
compute node on Summit has 6 GPU and 42 CPU, we are able to run 6 OpenMM tasks
in parallel which need 1 GPU and 1 CPU each. For optimal resource utilization,
we assign the remaining 36 CPU to 1 NAMD task. NAMD tasks run concurrently on
CPU with the OpenMM tasks running on GPU for \textit{heterogeneous parallelism}
(HP). In order to achieve the optimal processor utilization, CPU and GPU
computations must overlap as much as possible. We experimentally evaluate
heterogeneous parallelism in~\S\ref{ssection:expwfhybrid}.

\section{Performance Characterization}
\label{sec:performance}

\subsection{WF1: Ensemble Consensus Docking}\label{ssection:expwf1}

Compared to physics-based simulation methods, docking is a relatively
inexpensive computational process. To increase the reliability of docking
results, multiple docking protocols for the same ligand set and protease are
preferred over individual docking scores. WF1 uses OpenEye and Autodock-GPU.
The former is executed on x86 architectures (e.g., Frontera); the latter on
GPUs (e.g., Summit).

For each of the identified receptor sites, WF1 iterates through a list of
ligands and computes a docking score for that ligand-receptor pair. The score is
written to disk and is used as filter to identify ligands with favorable docking
results (score and pose). The docking call is executed as a Python function in
OpenEye, and as a self-contained task process in AutoDock-GPU. In both cases,
the RAPTOR framework (\S\ref{sssection:raptor}, Fig.\ref{sfig:raptor}) is used
for orchestration.

The duration of the docking computation depends on the type of CPU (OpenEye) or
GPU (AutoDock-GPU) used, and the computational requirements of each individual
receptor. We measure the docking time (seconds) and docking rate (docks/hr) of
three use cases: (1) production runs for NVBL-Medical Therapeutics campaigns;
and runs for largest achievable size on (2) Frontera and (3) Summit.
Table~\ref{tab:wf1} summarizes the parameterization and results of the
experiments we performed for each use case.

\begin{table*}
  \caption{WF1 use cases. For each use case, RAPTOR uses one pilot for each
  receptor, computing the docking score of a variable number of ligands to that
  receptor. OpenEye and AutoDock-GPU implement different docking algorithms and
  docking scores, resulting in different docking times and rates. However,
  resource utilization is $>=$90\% for all use cases.}
  \label{tab:wf1}
  \centering
  \begin{tabular}{ c  
                   l  
                   l  
                   r  
                   r  
                   r  
                   r  
                   r  
                   r  
                   r  
                   r  
                   r  
                   r} 
  \toprule
  \textbf{Use}                               &
  \multirow{2}{*}{\textbf{Platform}}         &
  \multirow{2}{*}{\textbf{Application}}      &
  \multirow{2}{*}{\textbf{Nodes}}            &
  \multirow{2}{*}{\textbf{Pilots}}           &
  \textbf{Ligands}                           &
  \multirow{2}{*}{\textbf{Utilization}}      &
  \multicolumn{3}{c}{\textbf{Docking Time [sec]}}  &
  \multicolumn{3}{c}{\textbf{Docking Rate [$\times10^6$ docks/hr]}}  \\
  \cline{8-13}
  \textbf{Case} &
                &
                &
                &
                &
  \textbf{[$\times10^6$]}  &
                &
  \textbf{min}  &
  \textbf{max}  &
  \textbf{mean} &
  \textbf{min}  &
  \textbf{max}  &
  \textbf{mean} \\
  \midrule
  1             &  
  Frontera      &  
  OpenEye       &  
  128           &  
  31            &  
  205           &  
  89.6\%        &  
  0.1           &  
  3582.6        &  
  28.8          &  
  0.2           &  
  17.4          &  
  5.0           \\ 
  2             &  
  Frontera      &  
  OpenEye       &  
  3850         &  
  1             &  
  125           &  
  95.5\%        &  
  0.1           &  
  833.1         &  
  25.1          &  
  16.0          &  
  27.5          &  
  19.1          \\ 
  3             &  
  Summit        &  
  AutoDock-GPU  &  
  1000         &  
  1             &  
  57            &  
  $\approx$95\% &  
  0.1           &  
  263.9         &  
  36.2          &  
  10.9          &  
  11.3          &  
  11.1          \\ 
  \bottomrule
  \end{tabular}
\end{table*}

WF1 assigns one pilot for each receptor to which a set of ligands will be
docked. Within each pilot, one master task is executed for every $\approx$100
nodes. Each master iterates at different offsets through the ligands database,
using pre-computed data offsets for faster access, and generating the docking
requests to be distributed to the worker tasks. Each worker runs on one node,
executing docking requests across the CPU cores/GPUs of that node.


\subsubsection{Use Case 1}

We assigned each of the 31 receptors to a single pilot \textemdash{} that is
an independent job submitted to the batch-queue. Due to the different
batch-queue waiting times, at most 13 concurrent pilots executed concurrently.
With 13-way pilot concurrency, the peak throughput was $\approx17.4\times10^6$
docks/hr. To keep an acceptable load on Frontera's shared filesystem, only 34
of the 56 cores available were used.

\begin{figure}[H]
  \up\up
  \centering
  \subfloat[][]{
      \includegraphics[width=0.23\textwidth]{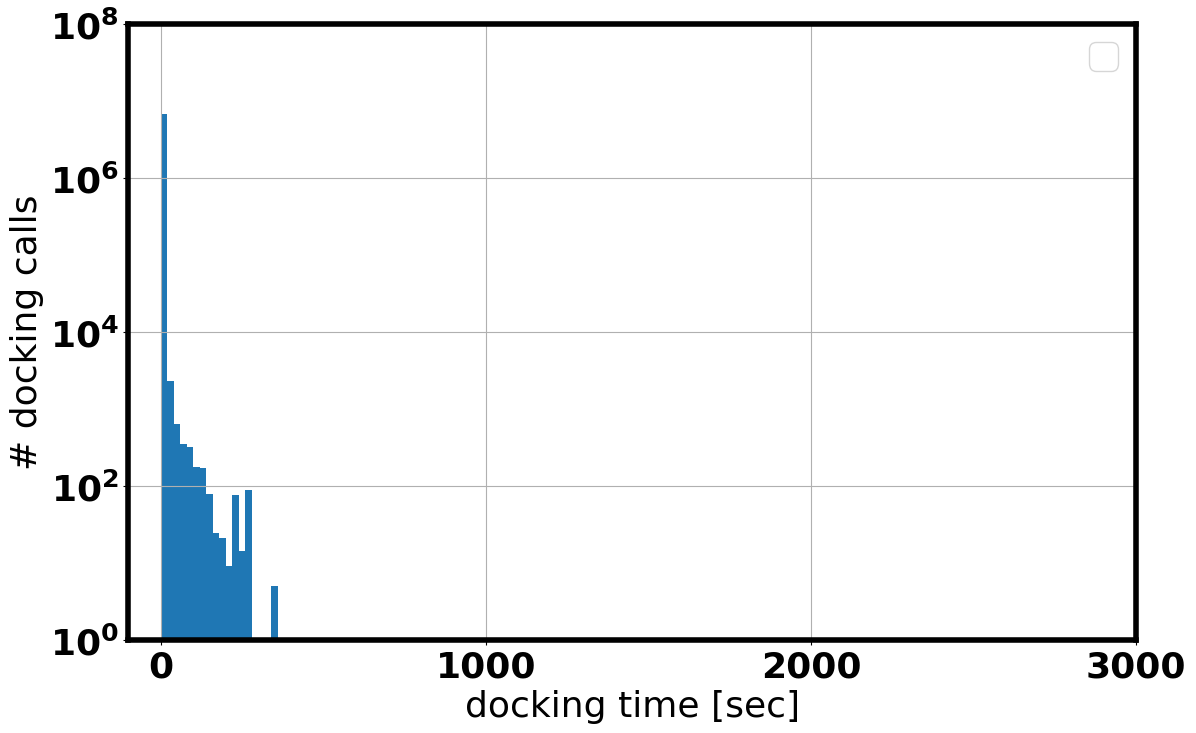}
      \label{sfig:wf1_uc1_durations_short}}
  \hfill
  \subfloat[][]{
      \includegraphics[width=0.23\textwidth]{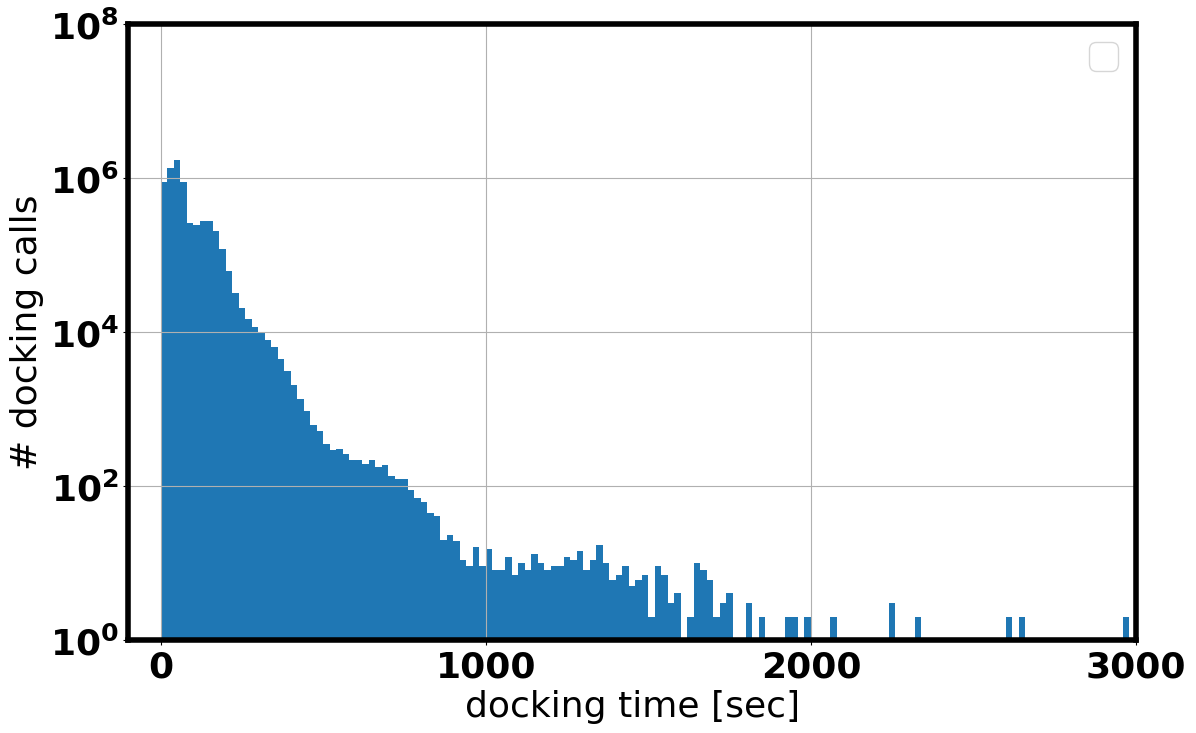}
      \label{sfig:wf1_uc1_durations_long}}
  \caption{WF1, Use Case 1: Distribution of docking runtimes for the receptor
           with the (a) shortest and (b) longest average docking time out of the
           31 receptors analyzed. The distributions of all 31 receptors have a
           long tail.}
  \up
  \label{fig:wf1_uc1_durations}
\end{figure}

Figs.~\ref{sfig:wf1_uc1_durations_short} and~\ref{sfig:wf1_uc1_durations_long}
show the distribution of docking times for receptors with the shortest and
longest average docking time, using the \texttt{Orderable-zinc-db-enaHLL}
ligand database. All receptors are characterized by long-tailed docking time
distributions. Across the 31 receptors, the min/max/mean docking times are
0.1/3582.6/28.8 seconds (Tab.~\ref{tab:wf1}), posing a challenge to
scalability due to the communication and coordination overheads. The long tail
distributions necessitates load balancing across available workers to maximize
resource utilization and minimize overall execution time.



We addressed load balancing by: (i) communicating tasks in bulk so as to limit
the communication frequency and therefore overhead; (ii) using multiple master
processes to limit the number of workers served by each master, avoiding
bottlenecks; (iii) using multiple concurrent pilots to partition the docking
computations of the set of ligands.

\begin{figure}[H]
  \up\up
  \centering
  \subfloat[][]{
    \includegraphics[width=0.23\textwidth]{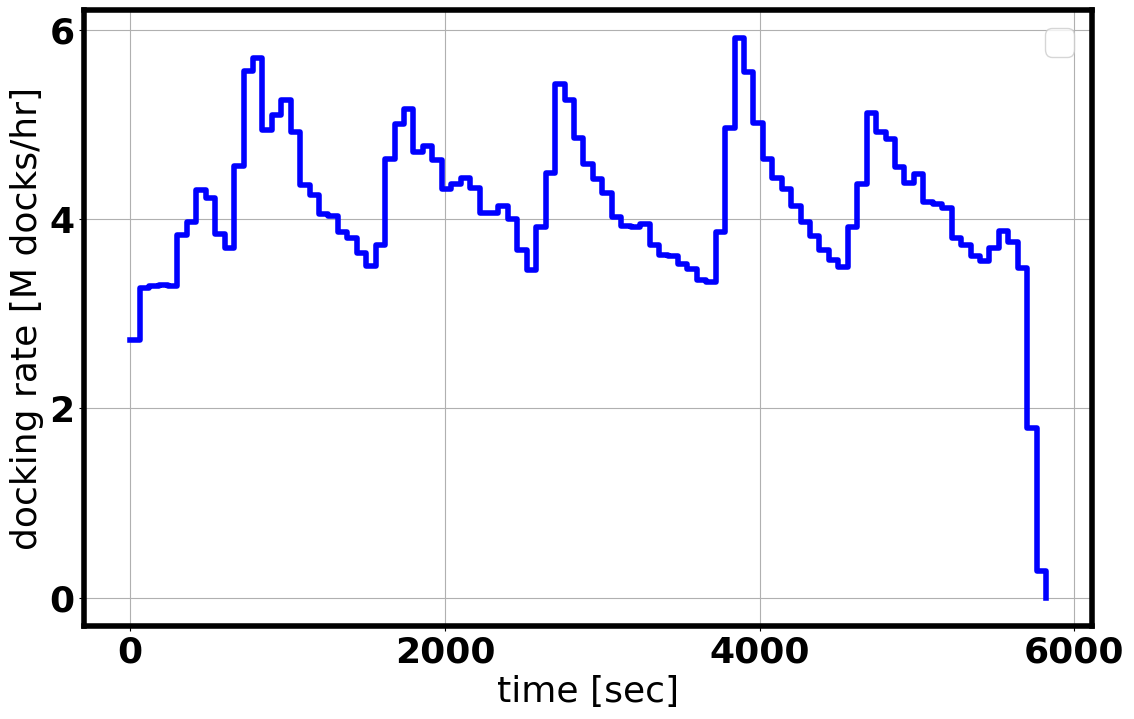}
    \label{sfig:wf1_uc1_rate_short}}
  \hfill
  \subfloat[][]{
    \includegraphics[width=0.23\textwidth]{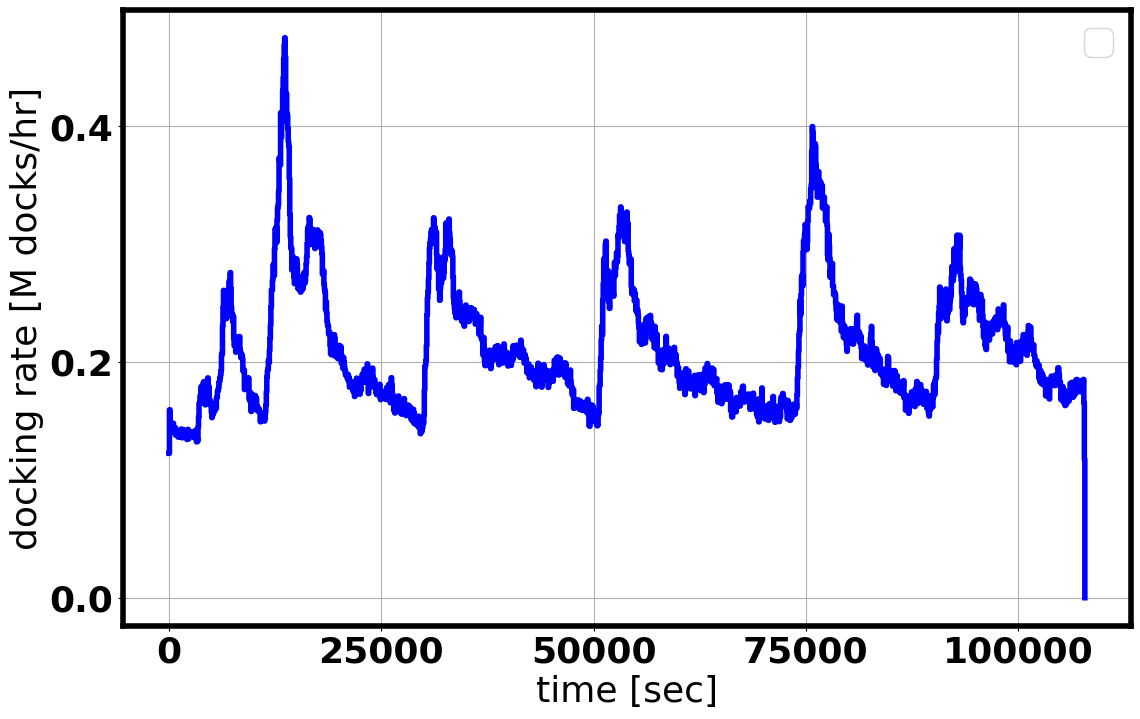}
    \label{sfig:wf1_uc1_rate_long}}
  \up
  \caption{WF1, Use Case 1: Docking rates for the receptor with  (a) shortest,
           and (b) longest average docking time.}
  \label{fig:wf1_uc1_rate}
  \up\up
\end{figure}

Figs.~\ref{sfig:wf1_uc1_rate_short} and~\ref{sfig:wf1_uc1_rate_long} show the
docking rates for the pilots depicted in
Figs.~\ref{sfig:wf1_uc1_durations_short} and~\ref{sfig:wf1_uc1_durations_long}
respectively. As with dock time distributions, the docking rate behavior is
similar across receptors. 
It seems likely that rate fluctuations depend on the interplay of machine
performance, pilot size, and specific properties of the ligands being
docked, and the target receptor. We measure a min/max docking rate of
$0.2/17.4\times10^6$ docks/hr with a mean of $5\times10^6$ docks/hr
(Tab.~\ref{tab:wf1}).

\subsubsection{Use Case 2}
\label{sssec:wf1-uc2}
\begin{figure}
  \up\up
  \centering
  \subfloat[][]{
    \includegraphics[width=0.23\textwidth]{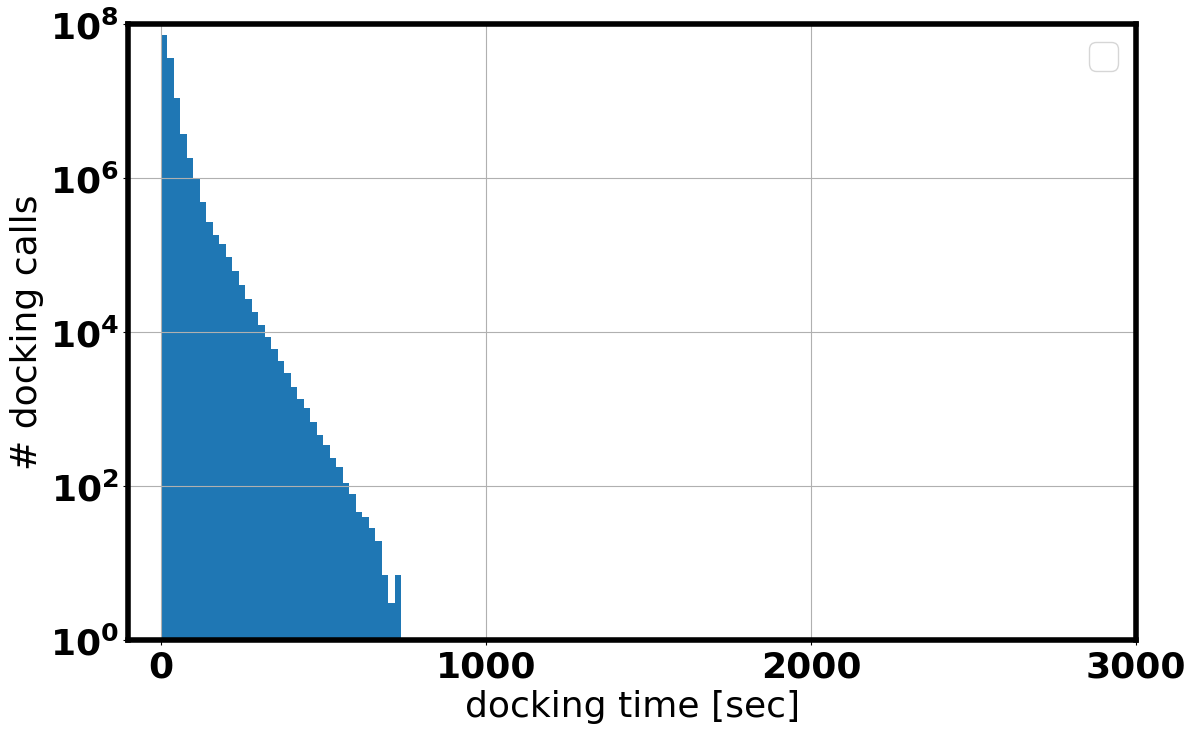}
    \label{sfig:wf1_uc1_time_hero}}
  \hfill
  \subfloat[][]{
    \includegraphics[width=0.23\textwidth]{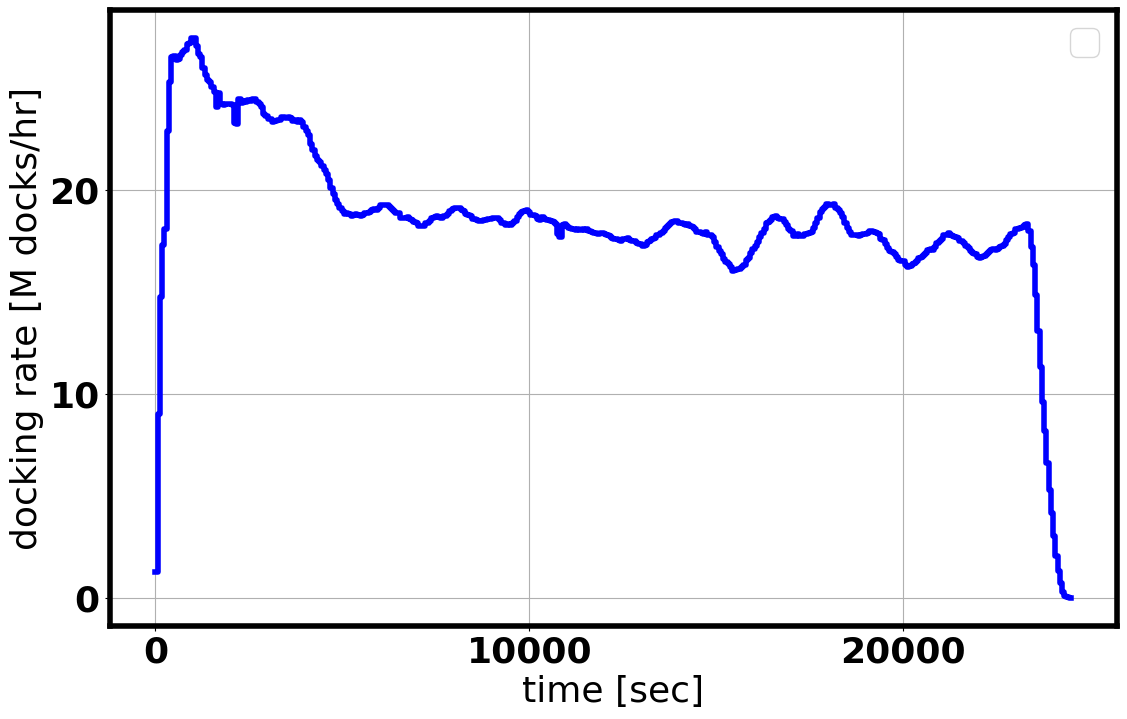}
    \label{sfig:wf1_uc1_rate_hero}}
  \up
  \caption{WF1, Use Case 2: (a) Distribution of docking time and (b) docking
           rate for a single receptor and $125\times10^6$ ligands. Executed with
           31 masters, each using $\approx$128 compute nodes/4352 cores on
           Frontera.}
  \label{fig:wf1_uc1_hero}
  \up\up\up
\end{figure}

Fig.~\ref{sfig:wf1_uc1_time_hero} shows the distribution of docking times of
approximately $125\times10^6$ ligands from the
\texttt{mcule-ultimate-200204-VJL} library \amnote{added, and the 125M is the
full library}\jhanote{Thank you. Could you do the same in Use Case 1 and 3
please?} to a single receptor using OpenEye on Frontera. The distribution has
a min/max of 0.1/833.1 seconds and a mean of 25.1 seconds
(Tab.~\ref{tab:wf1}).

Fig.~\ref{sfig:wf1_uc1_rate_hero} shows the docking rate for a single pilot
with 3850 compute nodes and 130,900 cores. Compared to Use Case 1, the rate
does not fluctuate over time. After peaking at $\approx27.5\times10^6$
docks/hr, the rate stabilizes at $\approx18\times10^6$ docks/hr until the end
of the execution (Tab.~\ref{tab:wf1}). Note that this run was terminated by
Frontera's batch system at the end of the given walltime, thus
Fig.~\ref{sfig:wf1_uc1_rate_hero} does not show the ``cool down'' phase.

Use Case 2 reached 95.5\% core utilization but, as mentioned, we were able to
utilize only 34 of the 56 node cores due to filesystem performance limitations.

\subsubsection{Use Case 3}
\label{sssec:wf1-uc3}

\begin{figure}
  \up\up
  \centering
  \subfloat[][]{
    \includegraphics[width=0.23\textwidth]{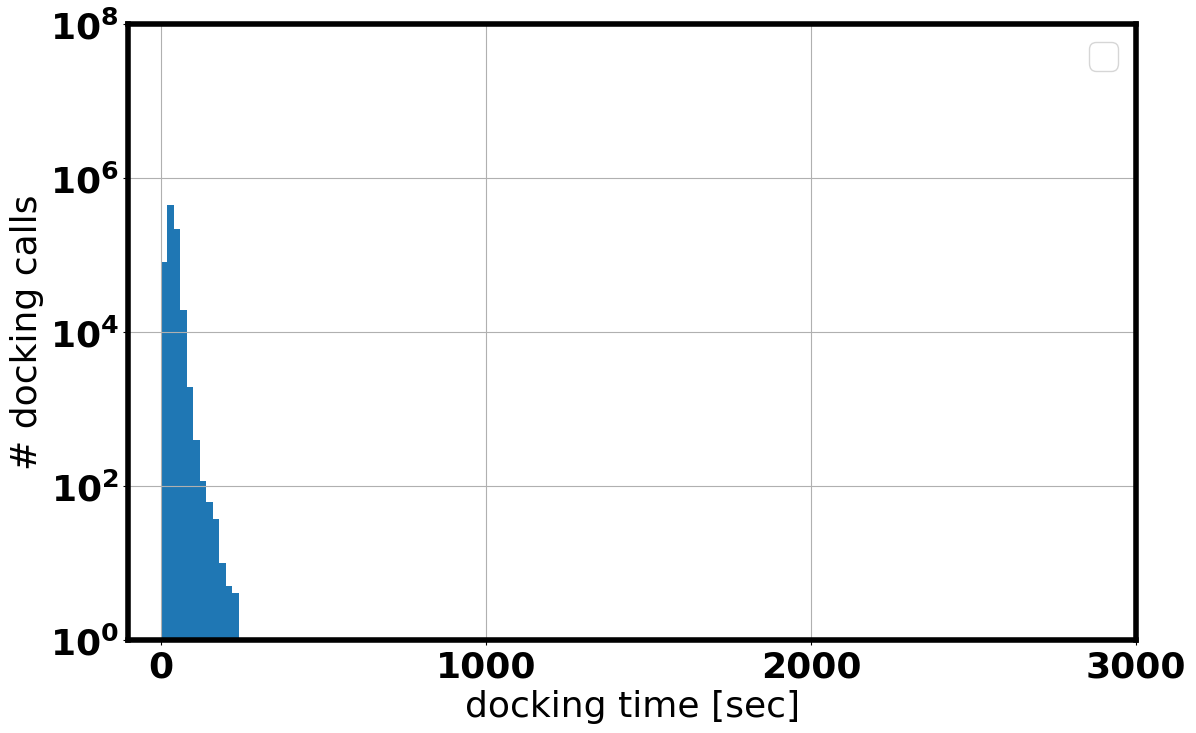}
    \label{sfig:wf1_uc1_time_summit_hero}}
  \hfill
  \subfloat[][]{
    \includegraphics[width=0.23\textwidth]{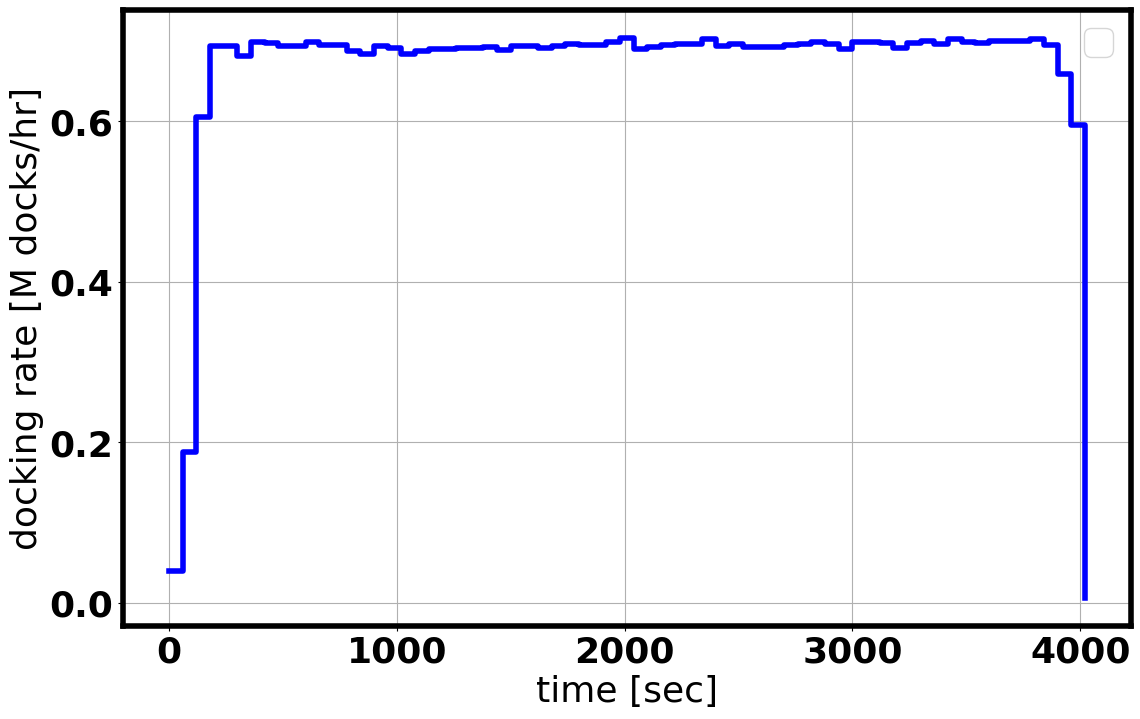}
    \label{sfig:wf1_uc1_rate_summit_hero}}
  \up
  \caption{WF1, Use Case 3: (a) Distribution of docking time and (b) docking
           rate for a single receptor and $57\times10^6$ ligands. Concurrently
           executed on Summit with a 6000 GPUs pilot.}
  \up\up\up
  \label{fig:wf1_uc1_summit_hero}
\end{figure}

Figure~\ref{sfig:wf1_uc1_time_summit_hero} shows the distribution of the docking
times of $\approx57\times10^6$ ligands from the \texttt{mcule-ultimate-200204-VJL}
database to a single receptor using AutoDock-GPU
on Summit. The distribution has a min/max/mean of 0.1/263.9/36.2 seconds
(Tab.~\ref{tab:wf1}). The max docking time is shorter than both Use Case 1,
Fig.~\ref{sfig:wf1_uc1_time_hero} and Use Case 2,
Fig.~\ref{fig:wf1_uc1_durations}, but the mean is longer. As observed, those
differences are likely due to specific properties of the docked ligands and the
target receptor.

Fig.~\ref{sfig:wf1_uc1_rate_summit_hero} shows the docking rate for a single
pilot with 1000 compute nodes, i.e., 6000 GPUs. Different from Use Case 1 and
2, the rate peaks very rapidly at $\approx11\times10^6$ docks/hr and maintains
that steady rate until the end of the execution. Cool down phase is also very
rapid. We do not have enough data to explain the observed sustained 
dock rate. As with Use Case 2, we assume an
interplay between the scoring function and its implementation in AutoDock-GPU
and specific features of the $57\times10^6$ docked ligands.

Different from OpenEye on Frontera, AutoDock-GPU bundles 16 ligands into one GPU
computation in order to efficiently use the GPU memory, reaching an average
docking rate of $11.1\times10^6$ docks/hr (Tab.~\ref{tab:wf1}). Currently, our
profiling capabilities allow us to measure GPU utilization with 5\% relative
error. Based on our profiling, we utilized between 93 and 98\% of the available
GPU resources.

\subsection{WF2: ML-Driven Enhanced Sampling}\label{ssection:expwf2}

WF2 is an iterative pipeline composed of 4 stages. After the first iteration of
the 4 stages is completed, if outliers were found, the next iteration starts
simulating those outliers; otherwise the simulation continues from where it
stopped in the previous iteration. The pipeline stops after a predefined number
of iterations.

We measured RCT overhead and resource utilization of WF2 to identify
performance bottlenecks. We define RCT overhead as the time spent not
executing at least one task. For example, the time spent bootstrapping
environments before tasks execution, communicating between EnTK and RabbitMQ
(RMQ), or between EnTK and RP while workloads wait to execute. Resource
utilization is the percentage of time when resources (CPUs and GPUs) are busy
executing tasks.

\begin{figure}[h]
  \up
  \centering
  \includegraphics[width=\linewidth]{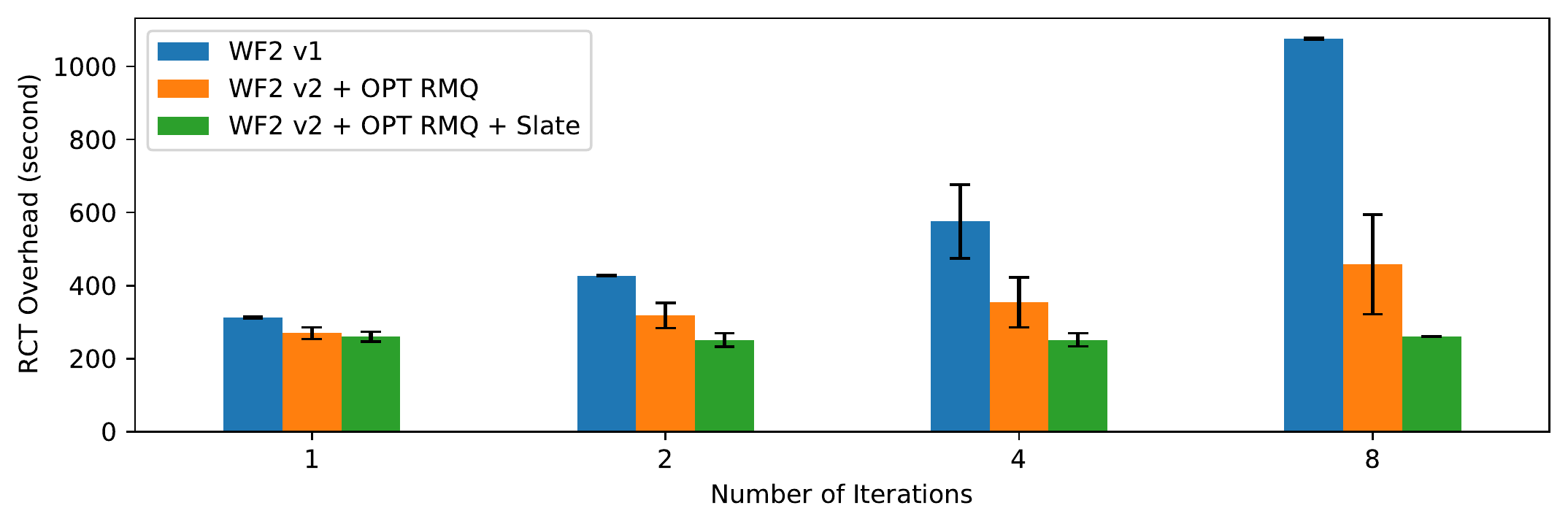}
  \up\up\up\up
  \caption{RCT overhead reduction with improved WF2, EnTK and RabbitMQ.}
  \up
  \label{fig:ml_driven_enhanced_sampling_rct_ovh}
\end{figure}

The blue bars in Fig.~\ref{fig:ml_driven_enhanced_sampling_rct_ovh} show RCT
overheads for the first version of WF2 and how RCT overheads grew with the
number of iterations.  WF2 may require a variable number of iterations. Thus,
our goal was to reduce RCT overhead, and importantly, to make it invariant of
the number of iterations.

An initial analysis suggested multiple optimizations of WF2: some of these
involved improving the deep learning model and the outlier detection of
DeepDriveMD, others required improving RCT. For the latter, we improved the
communication protocol between EnTK and RMQ, and we reduced the communication
latency between EnTK and RMQ. We avoided sharing connections to RMQ among EnTK
threads, reduced the number of concurrent multiple connections and reused
communication channels whenever possible.

Fig.~\ref{fig:ml_driven_enhanced_sampling_rct_ovh} (orange) shows the combined
effects of improving DeepDriveMD and EnTK communication protocol. Overheads
were reduced by 57\% compared to
Fig.~\ref{fig:ml_driven_enhanced_sampling_rct_ovh} (blue) but they were still
growing with the number of iterations. We moved our RMQ server to Slate, a
container orchestration service offered by OLCF. That reduced the
communication latency between EnTK and RMQ, as shown in
Fig.~\ref{fig:ml_driven_enhanced_sampling_rct_ovh} (green). This allowed RCT
overheads to be invariant up to 8 WF2 iterations.

\begin{figure}[h]
  \centering
  \includegraphics[width=\linewidth]{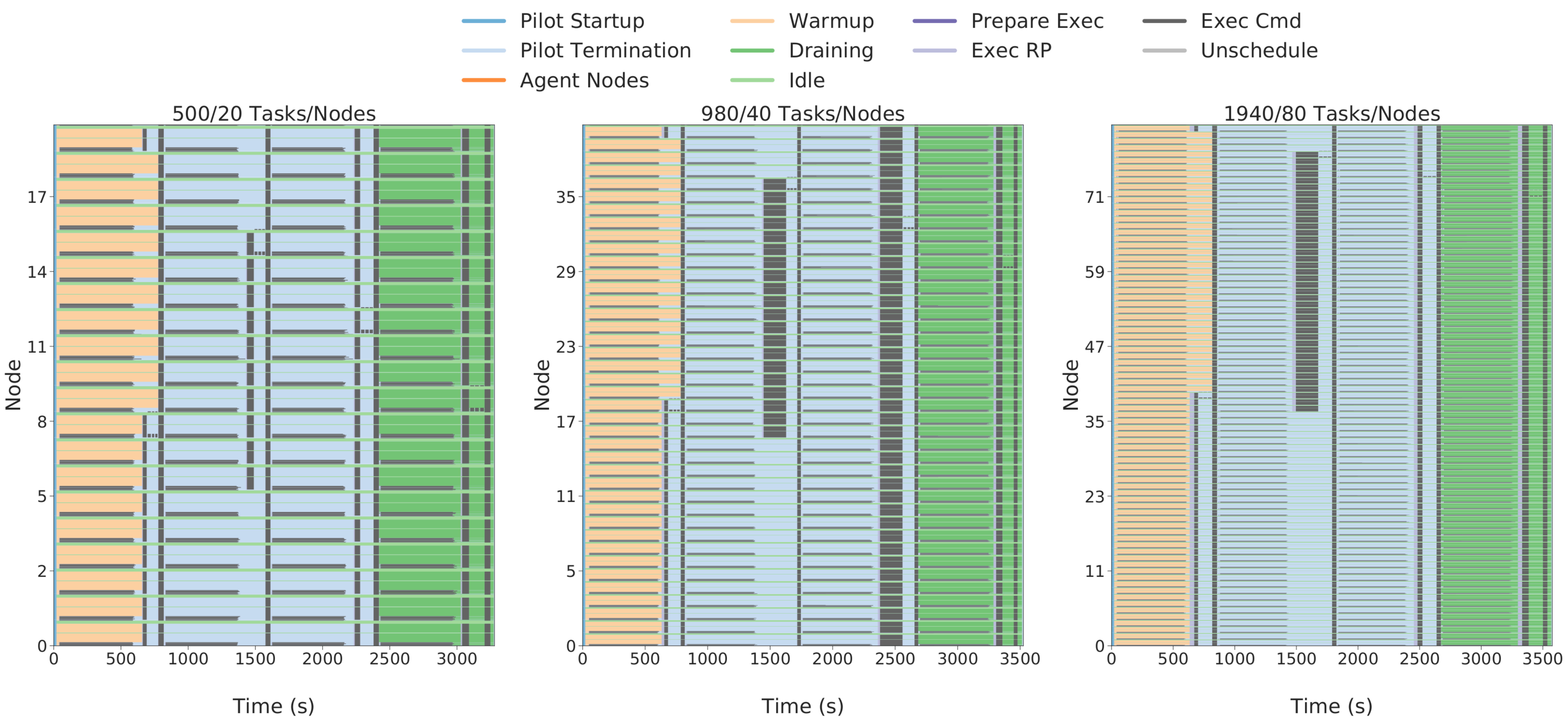}
  \up\up
  \caption{WF2 Resource Utilization: 20 nodes (left), 40 nodes (center), and 80
           nodes (right).}
  \up\up
  \label{fig:wf2-ru}
\end{figure}

Fig.~\ref{fig:wf2-ru} depicts resource utilization for different (internal) RCT
states as a time-series. ``Yellow, light blue and Green space'' represents
unused resources; ``dark'' represents resource usage. Fig.~\ref{fig:wf2-ru}
shows resource utilization of WF2, when executing four pipeline iterations on
Summit with 20, 40, and 80 compute nodes. Note that most of the unused resources
are CPU cores that are not needed by WF2. Overall, we measured 91\%, 91\%, 89\%
GPU utilization respectively. Across scales, Fig.~\ref{fig:wf2-ru} shows
differences in the execution time of some of the pipeline stages but no relevant
increase of the time spent without executing at least one task.


\subsection{Hybrid Workflows Using Heterogeneous Tasks}
\label{ssection:expwfhybrid}

WF3 and WF4 are computationally intensive methods that cost several orders of
magnitude more node-hours per ligand than WF1~\cite{saadi2020impeccable}. As
discussed in \S\ref{ssection:infwfhybrid}, WF3 and WF4 both compute binding
free energies, but have workloads comprised of distinct tasks: GPU-based
OpenMM, and CPU-based NAMD tasks respectively. Merging WF3 and WF4 into a
single hybrid workflow allowed us to improve resource utilization by employing
RP's unique capability of concurrently executing distinct tasks on CPU cores
and GPUs. We evaluated that capability by measuring: (i) RCT overhead (as
defined previously) as a function of scale; (ii) scalability as a function of
problem and resource size; and (iii) resource utilization.

\begin{figure}[h]
  \up
  \begin{center}
    \includegraphics[width=0.48\textwidth]{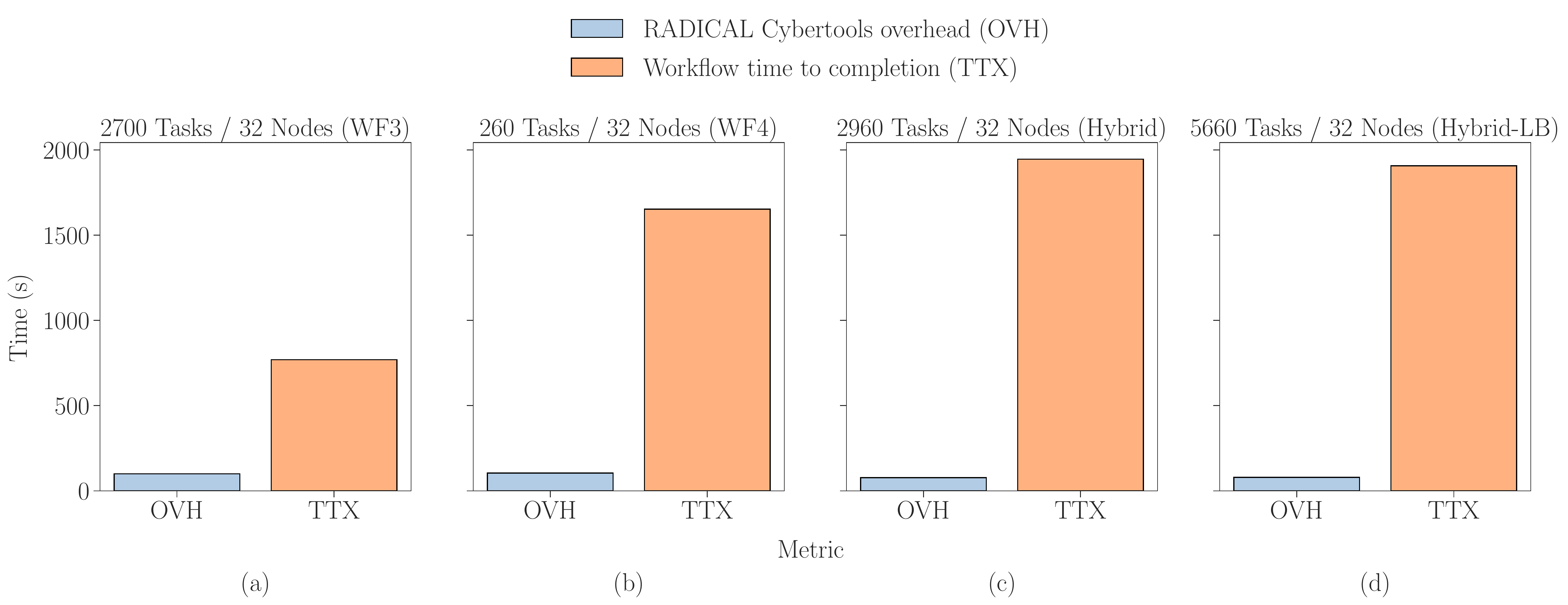}
  \end{center}
  \up\up\up\up
  \caption{RCT Overhead in Hybrid Workflows.}
  \up
  \label{fig:overhead_hybrid_base}
\end{figure}

Fig.~\ref{fig:overhead_hybrid_base} compares RCT overhead to workflow time to
completion (TTX) on 32 nodes for different tasks counts, representing
different production workflow configurations. TTX in
Fig.~\ref{fig:overhead_hybrid_base}(c) illustrates concurrent execution of GPU
and CPU tasks. The modest increase in TTX compared to
Fig.~\ref{fig:overhead_hybrid_base}(b) is likely due to interference from
sharing resources across tasks (Fig.~\ref{fig:hybrid}), and some scheduler
inefficiency. A careful evaluation and optimization will form the basis of
further investigation. Fig.~\ref{fig:overhead_hybrid_base}(d) plots the TTX
for the \emph{Hybrid-LB} scenario when the number of WF3 and WF4 tasks are
selected to ensure optimal resource utilization. The number of WF3 tasks
completed in Fig.~\ref{fig:overhead_hybrid_base}(d) is twice the number of WF3
tasks completed in Fig.~\ref{fig:overhead_hybrid_base}(c), with no discernible
increase in TTX.

\begin{figure}[h]
  \begin{center}
    \includegraphics[width=0.48\textwidth]{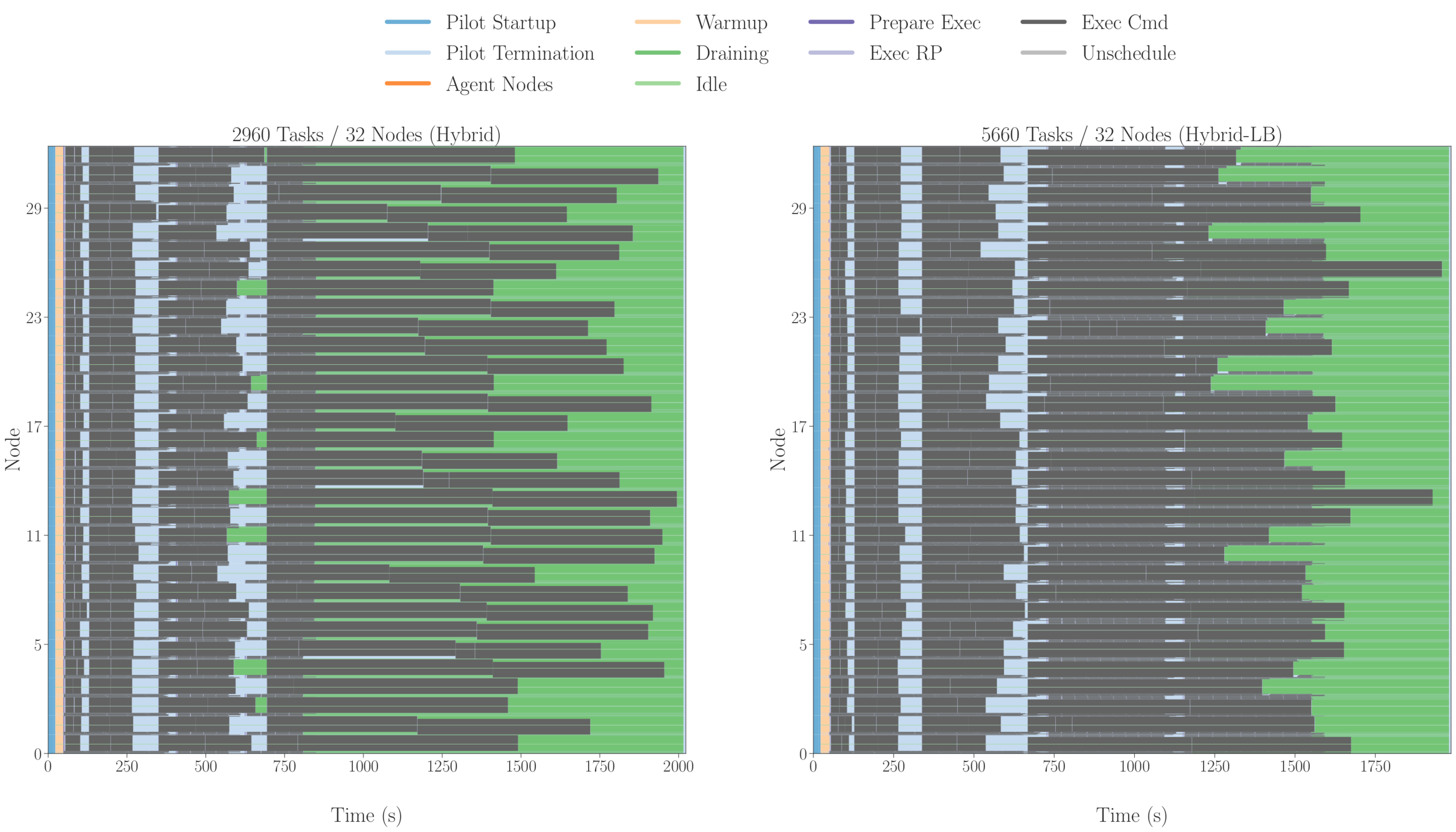}
  \end{center}
  \up\up
  \caption{RCT Resource Utilization in Hybrid Workflows.}
  \up\up
  \label{fig:resource_utilization_hybrid_base}
\end{figure}

Fig.~\ref{fig:resource_utilization_hybrid_base} depicts RCT resource
utilization for the configurations of Fig.~\ref{fig:overhead_hybrid_base}(c)
and Fig.~\ref{fig:overhead_hybrid_base}(d). As with Fig.~\ref{fig:wf2-ru},
``green space'' represents unused resources; ``dark space'' represents
resource usage. WF3 and WF4 have 4 and 3 stages respectively, which can be
discerned from black bars. Fig.~\ref{fig:resource_utilization_hybrid_base}(b)
shows greater dark space and thus resource utilization than
Fig.~\ref{fig:resource_utilization_hybrid_base}(a), representing greater
overlap of tasks on GPUs and CPUs due to workload sizing. Both have higher
resource utilization than configurations in
Fig.~\ref{fig:overhead_hybrid_base}(a) and (b) due to concurrent CPU and GPU
usage.

\begin{figure}[h]
  \begin{center}
    \includegraphics[width=0.48\textwidth]{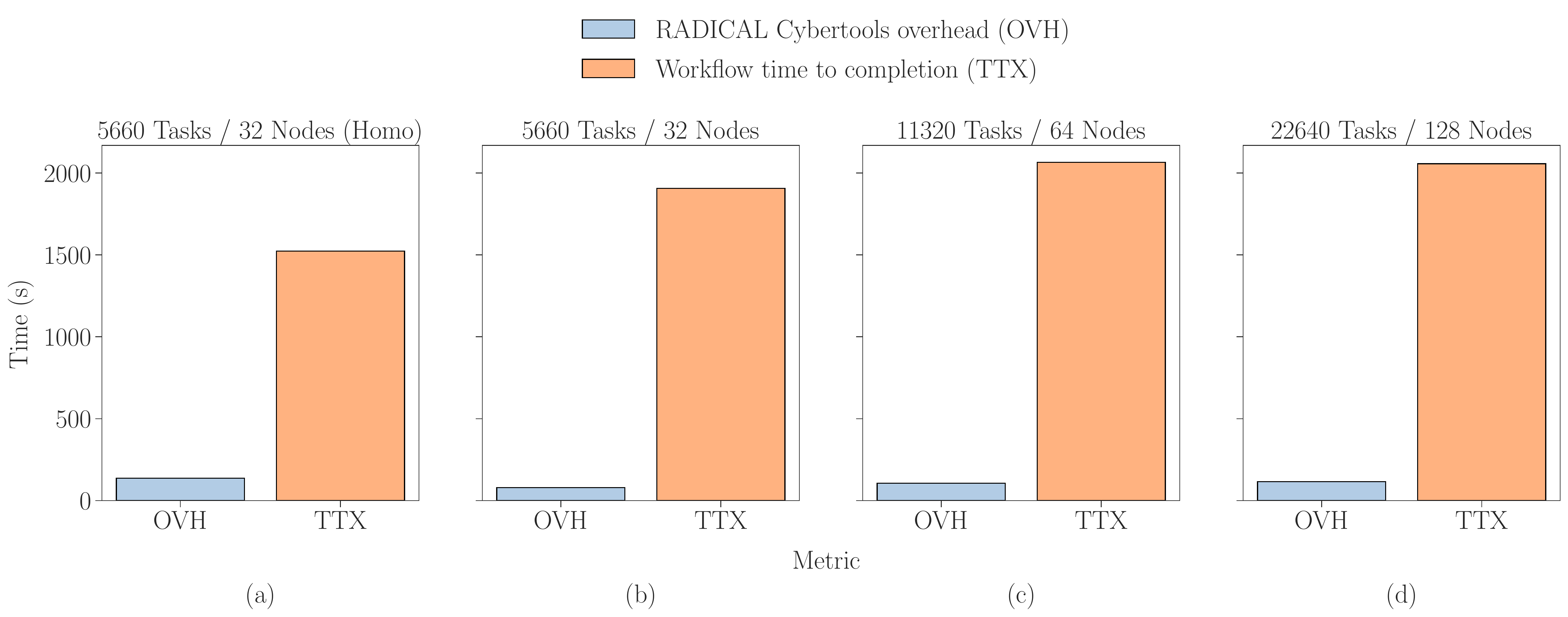}
  \end{center}
  \up\up\up
  \caption{RCT Overhead in Hybrid Workflows at Scale.}
  \up
  \up
  \label{fig:overhead_hybrid_scaling}
\end{figure}

Fig.~\ref{fig:overhead_hybrid_scaling} shows the scalability of hybrid
workflows with load balance enabled and up to 22640 tasks on 128 compute nodes
on Summit. The left two panels show the comparison between 5660 GPU tasks and
5660 heterogeneous tasks (5400 GPU tasks + 260 CPU tasks). Note that RCT
overhead is invariant between homogeneous and heterogeneous task placements,
and with proportionately increasing workloads and node counts (i.e., weak
scaling).

In Figs.~\ref{fig:overhead_hybrid_base} and~\ref{fig:overhead_hybrid_scaling},
RCT overhead varies from 3.8\% to 11.5\% of TTX but it should be noted that task
runtimes for these experiments are significantly shorter than those of
production runs.
RCT overhead arises from state transitions and data movements and is essentially
invariant of task runtimes, which are reproduced with fidelity in our
experiments. Thus, in production runs, RCT overhead is a significantly less
proportion of TTX.


\subsubsection*{Scaling WF3--4} Driven by science results
(\S~\ref{sec:result}), and that WF3 \& 4 are the ``slowest'' per
ligand~\cite{saadi2020impeccable}, we need to increase the number of nodes and
improve reliability across multiple platforms. We preview results for WF3;
experience with WF4 and hybrid WF3--4 execution will be reported subsequently.

We performed initial test runs using the multi-DVM execution mode described in
\S\ref{sssection:multidvm}, Fig.~\ref{sfig:multidvm}, and observed that
executions were stable with each DVM running on $< 50$ nodes and executing
$< 200$ tasks. Beyond that, we observed interruptions or connectivity losses
between executors and DVMs. Further investigation will establish the
causes of those limits and possible solutions for higher scalability.

\begin{figure}[h]
  \begin{center}
  \includegraphics[width=\linewidth]{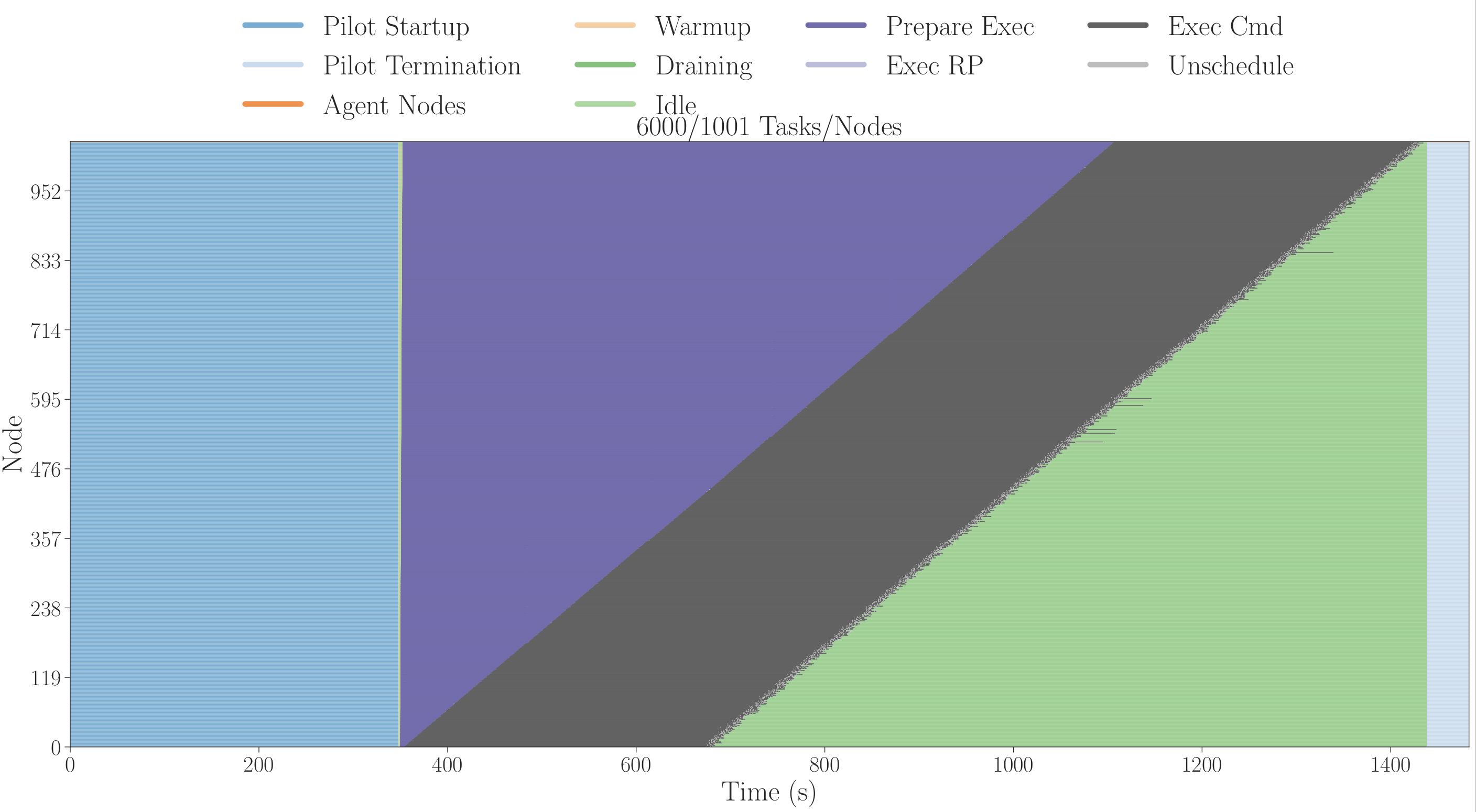}
  \end{center}
  \up\up
  \caption{RCT utilization for \wfcg using multi-DVM.}
  \up
  \label{fig:wf3cg_1000_utilization}
\end{figure}

We run \wfcg on 1000 compute nodes (+1 node for RTC), executing 6000 1-GPU tasks
on 32 concurrent DVMs. Each DVM spawned $\approx$32 nodes and executed up
to 192 tasks. Fig.~\ref{fig:wf3cg_1000_utilization} shows the utilization of the
available resources across different stages of the execution. The pilot startup
time (blue) is longer than when using a single
DVM~\cite{turilli2019characterizing}, mainly due to the 336 seconds spent on launching
DVMs which, currently, is a sequential process.
Each task requires time to prepare the execution (purple), which mainly includes
time for scheduling the task on a DVM, construct the execution command, and
process placement and launching by the DVM. The scheduling process takes longer
than with a single DVM as it requires to determine which DVM should be used.
Further, the construction of the execution command includes a 0.1s delay to let
DVM finalize the previous task launching. As each operation is done sequentially
per RP executor component, the 0.1s delay accounts for 600s alone.

As with the other WF3--4 experiments, we reduced task runtimes to limit resource
utilization while faithfully reproducing RCT overhead. In
Fig.~\ref{fig:wf3cg_1000_utilization}, Exec Cmd (task runtime) would be 10 times
longer for a production run. Thus, the overheads introduced by using multiple
DVMs would have a lesser impact on the overall resource utilization.


We also run \wfcg on 2000 compute nodes (+1 node for RCT), doubling task 
and DVM number compared to the run with 1000 nodes. At that scale, we
observed three main issues: (i) DVM startup failure;
(ii) an internal failure of PRRTE;
and (iii) lost DVM connectivity.
The majority of tasks were successfully completed (11802 out of 12000), but
those issues prevented RCT to gracefully handle their termination.

Given the current fragility of PRRTE/PMIx, we are investigating executing WF3
with the RP/Flux integration described in \S\ref{sssection:flux},
Fig.~\ref{sfig:flux}. Tab.~\ref{tab:wf3-flux} summarizes results for our
initial test. Performance are comparable to RCT using a single-dvm, reducing
the overheads measured with the multi-dvm implementation. We experienced no
failures, and are now working on deeper integration to further scale our
tests. If the current results hold at higher scales, we plan to use the
RP/Flux integration to run the WF3--4 pipeline in production on both Summit
and Lassen.

\begin{table*}
  \caption{WF3 use case. Test runs with RP/Flux integration
           (\S~\ref{sssection:flux}, Fig.~\ref{sfig:flux}).}
  \label{tab:wf3-flux}
  \centering
  \begin{tabular}{ l  
                   l  
                   c  
                   c  
                   c  
                   c  
                   c  
                   c} 
  \toprule
  \textbf{Use Case}                        &
  \textbf{Platform}                        &
  \textbf{\# Nodes}                        &
  \textbf{\# Tasks}                        &
  \textbf{\# Failed Tasks}                 &
  \textbf{Flux Resource Utilization}  &
  \textbf{Task Scheduling Rate}      &
  \textbf{Execution Time}              \\
  \midrule
  WF3           &  
  Lassen        &  
  128           &  
  512           &  
  0             &  
  88\%          &  
  14.21 t/s     &  
  6m             \\ 
  \bottomrule
  \end{tabular}
  \up\up
\end{table*}

\begin{table*}[h]
  \caption{HPC platforms used for the computational campaign. To manage the
  complexity arising from heterogeneity within and  across platforms, requires
  middleware abstractions and design.}
  \label{tab:hpc_platforms}
  \centering
  \begin{tabular}{llllccr}

      \toprule
      \textbf{HPC Platform}                          &
      \textbf{Facility}                              &
      \textbf{Batch}                                 &
      \multicolumn{2}{l}{\textbf{Node Architecture}} &
      \textbf{Workflows}                             &
      \textbf{Max \# nodes}                          \\

      \textbf{}                                      &
      \textbf{}                                      &
      \textbf{System}                                &
      \textbf{CPU}                                   &
      \textbf{GPU}                                   &
      \textbf{}                                      &
      \textbf{utilized}                              \\
      \midrule

      Summit                                         &
      OLCF                                           &
      LSF                                            &
      2 $\times$ POWER9 \hfill (22 cores)            &
      6 $\times$ Tesla V100                          &
      WF1-4                                          &
      2000                                          \\

      Lassen                                         &
      LLNL                                           &
      LSF                                            &
      2 $\times$ POWER9 (22 cores)                   &
      4 $\times$ Tesla V100                          &
      WF2,3                                          &
      128                                           \\

      Frontera                                       &
      TACC                                           &
      Slurm                                          &
      2 $\times$ x86\_64 \hfill (28 cores)           &
      ---                                            &
      WF1                                            &
      3850                                          \\

      Theta                                          &
      ALCF                                           &
      Cobalt                                         &
      1 $\times$ x86\_64 \hfill (64 cores)           &
      ---                                            &
      WF1                                            &
      256                                  \\

      SuperMUC-NG                                    &
      LRZ                                            &
      Slurm                                          &
      2 $\times$ x86\_64 \hfill (24 cores)           &
      ---                                            &
      WF3-4                                          &
      6000 (with failures)                          \\

      \bottomrule
  \end{tabular}
\end{table*}

\section{Scientific Results}
\label{sec:result}



The previous section characterized the performance of the scalable HPC and AI
infrastructure developed to support campaigns to advance COVID-19
therapeutics. Constituent workflows embody a diverse range of computational
characteristics. Tab.~\ref{tab:hpc_platforms} summarizes the heterogeneous
platforms utilized, and maps them to specific workflows supported. Put
together, the campaign has utilized 2.5$\times 10^6$ node-hours on these
platforms to support: (i) docking $\sim$10$^{11}$ ligands with a peak docking
rate of up to 40$\times 10^6$ docks/hr; (ii) thousands of AI-driven enhanced
sampling simulations; (iii) computed binding free energies on $\sim$10$^{5}$
ligand-protein complexes, including  10$^{4}$ concurrently. In addition to
``raw'' scale, individual workflow components demonstrate 100$\times$ to
1000$\times$ scientific improvement over traditional methods.

\begin{figure}
    \centering
    \includegraphics[width=0.33\textwidth]{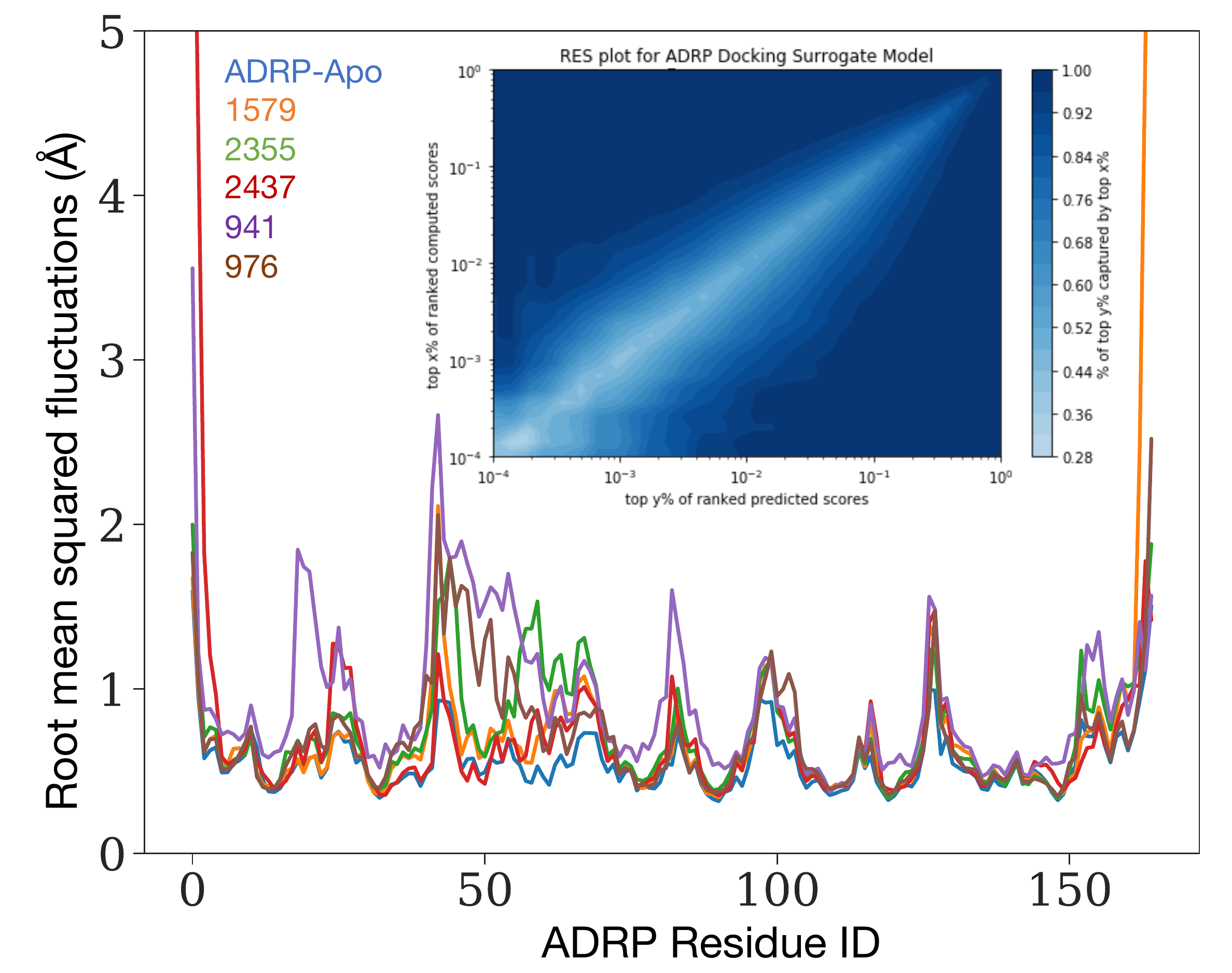}
    \up\up
    \caption{Root mean square fluctuation of ADRP protein and 5 bound ligands,
    illustrating that different ligands induce different conformational
    changes in ADRP. The inset highlights the RES based on the ML-based
    surrogate model, showing a clear trend in improving the number of
    downstream calculations needed for finding $n$ compounds in subsequent
    rounds of the iterative workflow. }
    \up\up
    \label{fig:adrp_rmsf}
\end{figure}

The scalable infrastructure provides unprecedented quantitative impact, but
also unique qualitative insight, that builds upon information multiple
workflows, an example of which we now preview. ML models used to predict
docking scores are inherently focused on predicting the ranking of small
molecules that potentially bind to and interact stably with the protein target
of interest (e.g., ADRP, as presented here). We utilize ML models to
accurately rank-order a library of ligands in terms of predicted ranked score,
using the regression enrichment surface (RES) technique to examine how well
the ML models act as a surrogate for the scoring
function\cite{clyde2020regression}. The RES plot (Fig. \ref{fig:adrp_rmsf}
inset) shows the surrogate model efficiency for detecting true top ranking
molecules given a fixed allocation of predicted hits. For instance, if the
computing budget allows $n$ number of downstream simulations for inferred
molecules of interest, the vertical line representing $n$ on the $x$-axis of
the plot shows the fraction of the real top scoring compound distribution
captured. Thus, the RES informs the number of top-scoring compounds needed to
adequately cover chemical space of ADRP-specific molecules.

Once a potential set of compounds predicted to bind to ADRP  are identified
using the RES technique, we characterize how stably they interact with ADRP.
We use DeepDriveMD (WF2) to study a small subset of compounds that potentially
interact with the primary binding site of ADRP\footnote{Although we studied
over 200 compounds, we present results from the top five compounds that
interact with ADRP stably during  O(100 ns) simulations}. To characterize the
stability, we chose to examine the root mean square fluctuation (RMSF)
analysis of the backbone C$^\alpha$ atoms for the \emph{apo}/ligand-free form
of the ADRP protein, which  shows decreased fluctuations in the location of
its backbone atoms compared to its \emph{holo}/ ligand-bound counterpart. The
ligand-bound protein undergoes higher RMSF fluctuations due to ligand-induced
conformational changes, with notable displacements of a few residues such as
Asp18-Val20, Asn54-Thr67, and His82. While these residues are not in direct
contact with the ligand, the binding pocket structural changes causes a ripple
effect on downstream residues in the protein.


\begin{figure}
    \centering
    \includegraphics[width=0.33\textwidth]{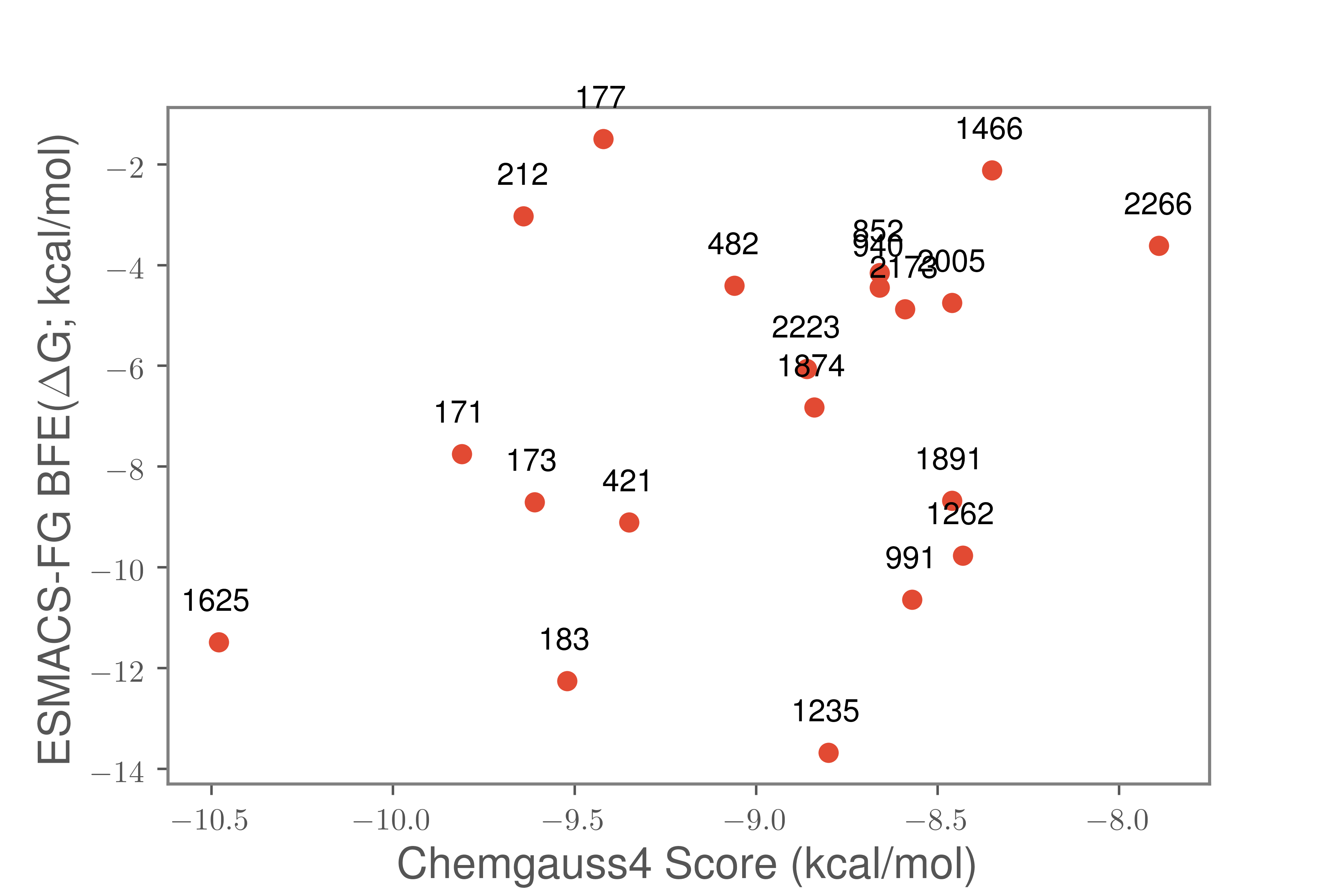}
    \up
    \caption{Correspondence between docking score (ChemGauss4/OpenEye)
    versus binding free energy using ESMACS-FG for ADRP. Although, the
    correlation is low, ESMACS-FG acts as an effective \emph{filter} for
    compounds that have high affinity to ADRP (e.g., compounds labeled 1625,
    183 and 1235.)}
    \up\up
    \label{fig:adrp_esmacs}
\end{figure}


We then use ESMACS (WF3) to predict which candidates bind tightly
(Fig.~\ref{fig:adrp_esmacs}). For instance, compound 1625, has high affinity
to ADRP from both ESMACS as well as docking (indicated by the ChemGauss4
docking score and the \esmacsfg binding free energy values), indicating
favorable interactions. TIES (WF4) is used to refine such interactions between
the protein and ligand:  we performed TIES on 19 compound transformations,
which entails mutating the original compound to new ligands with the goal of
improving binding affinity. We used this method to study the effect of
structural changes in a compound on their binding potency with ADRP.

The relative binding affinities ($\Delta\Delta G$) predicted by TIES for these
transformations fall between -0.55 to 4.62 kcal/mol. A positive value
indicates a diminished relative binding potency for the ``new'' compound,
whereas a negative value means that the transformation studied is favorable.
Twelve out of the 19 transformations studied have $\Delta\Delta G > +1$ which
suggests that they all correspond to unfavorable structural changes. The
remaining 7 have statistically zero value for $\Delta\Delta G$ which indicates
that the corresponding structural modifications do not affect the binding.
Both the favorable and unfavorable interactions provide insight into finding
compounds with high affinity for target proteins.


\section{Discussion}
\label{sec:discussion}





Multi-scale biophysics-based computational lead discovery is an important
strategy for accelerated drug development. In its current formulation and
practice however, it is too inefficient to explore drug compound libraries at
the scale of billions of molecules, even on the largest supercomputers. This
work demonstrates a path towards improvement by: (i) pairing ML components
with, and trained from physics-based components, and (ii) building the HPC and
AI infrastructure necessary to enable methodological
advances~\cite{saadi2020impeccable}. In doing so, it provides an enhanced drug
discovery pipeline for COVID-19~\textemdash{} a societal and intellectual
grand challenge.

This work is a harbinger of how the role of supercomputers is evolving, to
soon become increasingly important ``mere generators of data for powerful ML
models''. This will require the ability to generate data using computational
methods that have not historically been used at scale on supercomputers (e.g.,
WF1). Furthermore, scalable ML-driven methods are needed to improve
traditional physics-based approaches~\cite{geoffrey2019taxonomy-arxiv}.


Collectively, supercomputers will increasingly have to support campaigns with
diverse components, viz., physics-based simulations, data generation and
analysis, and ML/AI tasks. These individual workflows have different
computational characteristics and performance challenges. They encompass
high-throughput function calls, ensembles of MPI-based simulations, and
AI-driven HPC simulations. There are no ``turnkey solutions'' to support such
campaigns across multiple heterogeneous platforms, with the necessary
performance and scale to ensure the required throughput. This has necessitated
the design, development and iterative improvement of infrastructure to advance
therapeutics for COVID-19. The effectiveness and impact of the infrastructure
is evidenced by its use to sustain a campaign on multiple heterogeneous
platforms over a period of months to generate valuable scientific insight
(\S\ref{sec:result}).




\footnotesize{
\noindent {\bf Acknowledgements:}
Research was supported by the DOE Office of Science through the National
Virtual Biotechnology Laboratory; as part of the CANDLE project by the ECP
(17-SC-20-SC); UK MRC Medical Bioinformatics project (grant no. MR/L016311/1),
UKCOMES (grant no. EP/L00030X/1); EU H2020 CompBioMed2 Centre of Excellence
(grant no. 823712), and support from the UCL Provost.  Access to SuperMUC-NG
(LRZ) was made possible by a special COVID-19 allocation award from the Gauss
Centre for Supercomputing in Germany. Anda Trifan acknowledges support from
the United States Department of Energy through the Computational Sciences
Graduate Fellowship (DOE CSGF) under grant number: DE-SC0019323. We
acknowledge support and allocation from TACC and OLCF.





  \bibliographystyle{IEEEtran}
\bibliography{radical,references}

\begin{thebibliography}{10}
\providecommand{\url}[1]{#1}
\csname url@samestyle\endcsname
\providecommand{\newblock}{\relax}
\providecommand{\bibinfo}[2]{#2}
\providecommand{\BIBentrySTDinterwordspacing}{\spaceskip=0pt\relax}
\providecommand{\BIBentryALTinterwordstretchfactor}{4}
\providecommand{\BIBentryALTinterwordspacing}{\spaceskip=\fontdimen2\font plus
\BIBentryALTinterwordstretchfactor\fontdimen3\font minus
  \fontdimen4\font\relax}
\providecommand{\BIBforeignlanguage}[2]{{%
\expandafter\ifx\csname l@#1\endcsname\relax
\typeout{** WARNING: IEEEtran.bst: No hyphenation pattern has been}%
\typeout{** loaded for the language `#1'. Using the pattern for}%
\typeout{** the default language instead.}%
\else
\language=\csname l@#1\endcsname
\fi
#2}}
\providecommand{\BIBdecl}{\relax}
\BIBdecl

\bibitem{Bohacek_et_al:2010}
R.~S. Bohacek, C.~McMartin, and W.~C. Guida, ``The art and practice of
  structure-based drug design: A molecular modeling perspective,''
  \emph{Medicinal research reviews}, vol.~16, no.~1, pp. 3--50, 1996.

\bibitem{oetoolkit}
\BIBentryALTinterwordspacing
``Openeye toolkits 2019.oct.2,'' \emph{Open Eye Scientific}, 2019. [Online].
  Available: \url{http://www.eyesopen.com/}
\BIBentrySTDinterwordspacing

\bibitem{mcgann2003gaussian}
M.~R. Mcgann, H.~R. Almond, A.~Nicholls, J.~A. Grant, and F.~K. Brown,
  ``Gaussian docking functions,'' \emph{Biopolymers: Original Research on
  Biomolecules}, vol.~68, no.~1, pp. 76--90, 2003.

\bibitem{Maisuradze_2008}
\BIBentryALTinterwordspacing
G.~G. Maisuradze, A.~Liwo, and H.~A. Scheraga, ``Principal component analysis
  for protein folding dynamics,'' \emph{Journal of Molecular Biology}, vol.
  385, no.~1, pp. 312 -- 329, 2009. [Online]. Available:
  \url{http://www.sciencedirect.com/science/article/pii/S0022283608012886}
\BIBentrySTDinterwordspacing

\bibitem{Bhowmik_2018}
\BIBentryALTinterwordspacing
D.~Bhowmik, S.~Gao, M.~T. Young, and A.~Ramanathan, ``Deep clustering of
  protein folding simulations,'' \emph{BMC Bioinformatics}, vol.~19, no.~18, p.
  484, 2018. [Online]. Available:
  \url{https://doi.org/10.1186/s12859-018-2507-5}
\BIBentrySTDinterwordspacing

\bibitem{lee2019deepdrivemd}
H.~Lee, M.~Turilli, S.~Jha, D.~Bhowmik, H.~Ma, and A.~Ramanathan,
  ``Deepdrivemd: Deep-learning driven adaptive molecular simulations for
  protein folding,'' in \emph{2019 IEEE/ACM Third Workshop on Deep Learning on
  Supercomputers (DLS)}.\hskip 1em plus 0.5em minus 0.4em\relax IEEE, 2019, pp.
  12--19.

\bibitem{brd4}
S.~Wan, A.~P. Bhati, S.~J. Zasada, I.~Wall, D.~Green, P.~Bamborough, and P.~V.
  Coveney, ``Rapid and reliable binding affinity prediction of bromodomain
  inhibitors: A computational study,'' \emph{Journal of Chemical Theory and
  Computation}, vol.~13, no.~2, pp. 784--795, 2017.

\bibitem{ties}
\BIBentryALTinterwordspacing
A.~P. Bhati, S.~Wan, D.~W. Wright, and P.~V. Coveney, ``Rapid, accurate,
  precise, and reliable relative free energy prediction using ensemble based
  thermodynamic integration,'' \emph{Journal of Chemical Theory and
  Computation}, vol.~13, no.~1, pp. 210--222, 2017, pMID: 27997169. [Online].
  Available: \url{https://doi.org/10.1021/acs.jctc.6b00979}
\BIBentrySTDinterwordspacing

\bibitem{ti1}
\BIBentryALTinterwordspacing
T.~P. Straatsma, H.~J.~C. Berendsen, and J.~P.~M. Postma, ``Free energy of
  hydrophobic hydration: A molecular dynamics study of noble gases in water,''
  \emph{The Journal of Chemical Physics}, vol.~85, no.~11, pp. 6720--6727,
  1986. [Online]. Available: \url{https://doi.org/10.1063/1.451846}
\BIBentrySTDinterwordspacing

\bibitem{merzky2015saga}
A.~Merzky, O.~Weidner, and S.~Jha, ``{SAGA}: A standardized access layer to
  heterogeneous distributed computing infrastructure,'' \emph{Software-X}, vol.
  1-2, pp. 3--8, 2015.

\bibitem{merzky2018using}
A.~Merzky, M.~Turilli, M.~Maldonado, M.~Santcroos, and S.~Jha, ``Using pilot
  systems to execute many task workloads on supercomputers,'' in \emph{Workshop
  on Job Scheduling Strategies for Parallel Processing}.\hskip 1em plus 0.5em
  minus 0.4em\relax Springer, 2018, pp. 61--82.

\bibitem{balasubramanian2018harnessing}
V.~Balasubramanian, M.~Turilli, W.~Hu, M.~Lefebvre, W.~Lei, R.~Modrak,
  G.~Cervone, J.~Tromp, and S.~Jha, ``Harnessing the power of many: Extensible
  toolkit for scalable ensemble applications,'' in \emph{2018 IEEE
  International Parallel and Distributed Processing Symposium (IPDPS)}.\hskip
  1em plus 0.5em minus 0.4em\relax IEEE, 2018, pp. 536--545.

\bibitem{turilli2018comprehensive}
\BIBentryALTinterwordspacing
M.~Turilli, M.~Santcroos, and S.~Jha, ``A comprehensive perspective on
  pilot-job systems,'' \emph{ACM Comput. Surv.}, vol.~51, no.~2, pp.
  43:1--43:32, Apr. 2018. [Online]. Available:
  \url{http://doi.acm.org/10.1145/3177851}
\BIBentrySTDinterwordspacing

\bibitem{quintero2019ibm}
D.~Quintero, M.~Gomez~Gonzalez, A.~Y. Hussein, and J.-F. Myklebust, \emph{{IBM}
  High-Performance Computing Insights with {IBM} Power System AC922 Clustered
  Solution}.\hskip 1em plus 0.5em minus 0.4em\relax IBM Redbooks, 2019.

\bibitem{castain2018pmix}
\BIBentryALTinterwordspacing
R.~H. Castain, J.~Hursey, A.~Bouteiller, and D.~Solt, ``{PMIx}: Process
  management for exascale environments,'' \emph{Parallel Computing}, vol.~79,
  pp. 9 -- 29, 2018. [Online]. Available:
  \url{http://www.sciencedirect.com/science/article/pii/S0167819118302424}
\BIBentrySTDinterwordspacing

\bibitem{turilli2019characterizing}
M.~Turilli, A.~Merzky, T.~Naughton, W.~Elwasif, and S.~Jha, ``Characterizing
  the performance of executing many-tasks on summit,'' \emph{IPDRM Workshop,
  SC19}, 2019, https://arxiv.org/abs/1909.03057.

\bibitem{ahn2014flux}
D.~H. Ahn, J.~Garlick, M.~Grondona, D.~Lipari, B.~Springmeyer, and M.~Schulz,
  ``Flux: a next-generation resource management framework for large hpc
  centers,'' in \emph{2014 43rd International Conference on Parallel Processing
  Workshops}.\hskip 1em plus 0.5em minus 0.4em\relax IEEE, 2014, pp. 9--17.

\bibitem{saadi2020impeccable}
A.~A. Saadi, D.~Alfe, Y.~Babuji, A.~Bhati, B.~Blaiszik, T.~Brettin, K.~Chard,
  R.~Chard, P.~Coveney, A.~Trifan \emph{et~al.}, ``Impeccable: Integrated
  modeling pipeline for covid cure by assessing better leads,'' \emph{arXiv
  preprint arXiv:2010.06574}, 2020.

\bibitem{clyde2020regression}
A.~Clyde, X.~Duan, and R.~Stevens, ``Regression enrichment surfaces: a simple
  analysis technique for virtual drug screening models,'' \emph{arXiv preprint
  arXiv:2006.01171}, 2020.

\bibitem{geoffrey2019taxonomy-arxiv}
G.~Fox and S.~Jha, ``Learning everywhere: A taxonomy for the integration of
  machine learning and simulations,'' \emph{IEEE eScience (2019)}, 2019, arXiv
  preprint arXiv:1909.13340.

\end{thebibliography}


\end{document}